\documentclass[fleqn,usenatbib]{mnras}

\usepackage{newtxtext,newtxmath}

\usepackage[T1]{fontenc}

\DeclareRobustCommand{\VAN}[3]{#2}
\let\VANthebibliography\thebibliography
\def\thebibliography{\DeclareRobustCommand{\VAN}[3]{##3}\VANthebibliography}

\usepackage{graphicx}	
\usepackage{amsmath}	

\title[The origin of scatter in the $L_{\rm X} (M_{\rm 500c})$ relation]{The origin of scatter in the X-ray luminosity -- halo mass relation of galaxy clusters}

\author[J. Braspenning et al.]{
Joey Braspenning$^{1,2}$\thanks{E-mail: jobraspenning@mpia.de},
Joop Schaye$^{2}$,
Annalisa Pillepich$^{1}$,
and Dylan Nelson$^{3}$
\\
$^{1}$Max-Planck-Institut für Astronomie, Königstuhl 17, D-69117 Heidelberg, Germany \\
$^{2}$Leiden Observatory, Leiden University, PO Box 9513, 2300 RA Leiden, the Netherlands\\
$^{3}$Universität Heidelberg, Zentrum für Astronomie, Institut für theoretische Astrophysik, Albert-Ueberle-Str. 2, D-69120 Heidelberg, Germany
}

\date{Accepted XXX. Received YYY; in original form ZZZ}

\pubyear{\the\year{}}

\begin{document}
\label{firstpage}
\pagerange{\pageref{firstpage}--\pageref{lastpage}}
\maketitle

\begin{abstract}
Galaxy groups and clusters are excellent probes of large-scale structure and are shaped by some of the most energetic physical processes in the Universe. They follow a tight scaling relation of X-ray luminosity with halo mass. However, predicting the dependence of the scatter in this relation on mass and redshift is challenging, due to the statistical requirement of large simulation volumes. Using the large volume cosmological hydrodynamical simulations for galaxy cluster physics FLAMINGO and TNG300+TNG-Cluster, we fit this relation and its scatter, focusing on $M_{\rm 500c}>10^{13}~\mathrm{M_\odot}$ and $z \leq 2$. We find qualitatively similar, but quantitatively different results for the two models. For the first time, we study ways to reduce the scatter using properties beyond X-ray luminosity, namely six ICM, six galaxy, and eleven dark matter halo properties. For both FLAMINGO and TNG300+TNG-Cluster, the gas fraction and thermal Sunyaev-Zel\'dovich (SZ) signal correlate strongest with X-ray scatter, reducing it by over 50\% when accounting for their partial correlations. Galaxy and halo properties correlate weakly with X-ray scatter, typically reducing it by 10-20\%. Our results are qualitatively robust across different FLAMINGO feedback variations, though the correlations weaken for stronger feedback and with increasing redshift. 
Differences between FLAMINGO and TNG300+TNG-Cluster are only apparent at the high-mass end -- where e.g. the galaxy stellar age correlates strongly for FLAMINGO, but not for TNG300+TNG-Cluster -- confirming robustness across physics implementations. We provide fitting formulas for the scatter and its corrections, for direct application to cosmological analyses and observational data.
\end{abstract}




\section{Introduction}

Galaxy groups and clusters are complex systems. Although the main observables correlate strongly with halo mass, even the most massive galaxy clusters are thought to have ejected a significant fraction of their gas during their evolution \citep{mitchell_baryonic_2022}. This violent past left imprints on the clusters as we see them today. For example, the X-ray luminosity at a fixed halo mass is correlated with having a cool-core in observations \citep{pratt_galaxy_2009}, and with halo concentration in simulations \citep{fujita_halo_2019}, but the observational dependence on the correlation with the dynamical state is debated \citep[e.g.][]{pratt_galaxy_2009, eckmiller_testing_2011}. Past merger and Active Galactic Nucleus (AGN) activity are also expected to affect the current cool-core status of a cluster \citep[e.g.][]{lehle_what_2025}. The past histories of galaxy groups and clusters can cause coherent deviations from the $z=0$ scaling relations. Such correlations between the scatter in different relations with halo mass are observed, for example, between the X-ray luminosity and gas temperature \citep{mantz_weighing_2016}.

From the point of view of cosmological structure formation, residual correlations between the past histories and current observables are expected.
In the current $\Lambda$CDM concordance model of cosmology \citep{planck_collaboration_planck_2014}
structure forms hierarchically, with massive haloes coming about from the mergers of smaller haloes, and hence forming later. As a result, the most massive haloes at the current time have a high merger rate \citep{lacey_merger_1993, stewart_galaxy_2009}. 
Hierarchical structure growth intuitively leads to the preconception that halo properties are mostly mass-dependent.
In this picture, haloes are statistically independent of their larger-scale environment. However, the situation is more complex. Halo assembly bias introduces the concept that massive haloes' clustering properties depend on not just the mass, but also the formation time \citep[e.g.][]{gao_age_2005, croton_halo_2007, zentner_galaxy_2014, liu_luminosity-dependent_2025}. This secondary dependence can be explained by haloes of different ages being at different stages of their evolution. 

Because the density of the Universe is higher at earlier times, haloes that form earlier have a higher concentration, and because they also have more time to relax, they are more concentrated at the present time \citep[e.g.][]{wechsler_concentrations_2002, zhao_accurate_2009, jeeson-daniel_correlation_2011, ludlow_massconcentrationredshift_2014, correa_accretion_2015}. At fixed halo mass, a strong correlation also exists between galaxy stellar mass and halo concentration, both in observations \citep[e.g.][]{zu_does_2021}, analytical models \citep[e.g.][]{bradshaw_physical_2020}, and simulations \citep[e.g.][]{matthee_origin_2017}. In hierarchical structure formation, the most massive objects form last, leading to them having the lowest concentration, which is indeed confirmed by observation and simulations up to the most massive galaxy clusters \citep{vikhlinin_chandra_2006, ettori_mass_2010, bahe_mock_2012}.

Cosmological models set specific expectations for the evolution of the halo mass relation with redshift. Extracting cluster halo masses from observations to compare with those expectations is challenging, and most commonly done through the use of mass-proxy scaling relations. The level of scatter of mass-proxy scaling relations directly translates into an uncertainty on parameters in cosmological parameter inference \citep{allen_cosmological_2011}. This is especially true for large all-sky surveys, where resolved information on a cluster-by-cluster basis is often unavailable. \citet{eckert_low-scatter_2020} point out how using metrics with reduced scatter yields significantly lower systematic errors on cosmological parameters. \citet{mulroy_locuss_2019} show how different scaling relations for the same clusters have varying amounts of scatter, and that weak correlations exist between morphology measurements and the scatter of some of these scaling relations. This is further reinforced by \citet{aljamal_mass_2025} who show, based on the outcome of cosmological hydrodynamical simulations, that different halo gas tracers result in different amounts of scatter in the halo mass estimate, with the scatter being strongly mass dependent.

As a means to reduce the scatter of cluster scaling relations, and hence of the halo mass estimate, previous observational work has often focused on creating subsamples of galaxy clusters with less scatter. A common distinction is between relaxed and non-relaxed clusters: the idea is that dynamically relaxed clusters are more self-similar, and hence scatter less around the median scaling relation. This approach includes weeding out recent mergers, in which, due to shocks and sloshing, the hot gas properties are not representative of the halo mass \citep[e.g.][]{poole_impact_2007}. As a population, dynamically-disturbed clusters also have a different luminosity function \citep{martinez_dynamical_2012}. However, selecting only relaxed clusters, with often stringent requirements on relaxedness, drastically reduces sample size: e.g. \citet{mantz_cosmology_2014} use only the 40 most relaxed clusters out of a sample of $>300$ for their cosmological analysis. On the other hand, \citet{ seppi_modelling_2025} find that comparing relaxed and unrelaxed systems from the TNG300 cosmological magnetohydrodynamical simulation of galaxies, and hence clusters, results in minimally different intrinsic scatters, and \citet{correa_magnus_luminosity-dependent_2025} find that the hot gas density and temperature profiles of these two classes are also only minimally different. 

There have been some tentative efforts to correct sample averages for the dynamical state. For example, \citet{shi_analytical_2016} introduce an analytical model to correct the sample average mass estimate for the non-thermal pressure contribution, and \citet{haggar_reconsidering_2024} study the correlation between the dynamical states of simulated clusters and mock observable metrics, finding sample average correlations. \citet{marini_impact_2025} show that in the Magneticum simulations (a suite of cosmological hydrodynamical simulations), there are not only correlations between X-ray luminosity and dynamical state, but also with certain other properties of the halo, such as super-massive black hole (SMBH) and stellar mass. Similarly, \citet{costello_luminosity-dependent_2025}, for the FLAMINGO simulations, find that the gas fraction of haloes is correlated with their SMBH mass. Yet, the halo-to-halo scatter is large. 

One of the scientific rationales for cosmological simulations is their ability to elucidate physical mechanisms underlying observable properties. While observations provide us only with a single snapshot in time, are subject to systematic uncertainties, and typically measure at most a few properties per object, simulations have access to the entire history of a halo and provide a nearly unlimited set of physical tracers. They furthermore allow us to go beyond analytical predictions and study residual correlations driven by non-linear physical processes, resulting in object-to-object scatter \citep[e.g.][]{matthee_origin_2017, martizzi_baryons_2020, pei_effectiveness_2024}, at the expense of the predictions being model dependent. 

In this work, we first aim at quantifying the mass-dependent scatter of the X-ray luminosity -- halo mass relation of galaxy clusters, and provide fitting functions for the scatter as a function of mass and redshift. We then focus on reducing the scatter of this scaling relation, and quantify for a large number of gas, stellar, SMBH and dark matter tracers how well they correlate with the scatter in the X-ray luminosity.

Thoroughly investigating the correlation among secondary properties requires a large sample of galaxy clusters. Because of their cosmological rarity, this was until recently impossible in either observations or cosmological hydrodynamical simulations that include galaxy astrophysics. The FLAMINGO cosmological hydrodynamical simulations change this by providing a large sample of galaxy groups ($> 2,000,000$) and clusters ($> 100,000$) in their full cosmological context, all the way to $z=0$ \citep{schaye_flamingo_2023, kugel_flamingo_2023}.

In this work, we use the outcome of the FLAMINGO simulations as well as of the TNG300 \citep{nelson_first_2018, springel_first_2018, pillepich_first_2018, naiman_first_2018, marinacci_first_2018} and TNG-Cluster \citep{nelson_introducing_2024} simulations (hereafter TNG300+TNG-Cluster), to check for possible effects of different galaxy-physics choices and implementations. From FLAMINGO we use the flagship $(2.8~\mathrm{Gpc})^3$ simulation, as well as the $(1.0~\mathrm{Gpc})^3$ baryonic feedback variations. TNG300 is the $(\approx 300~\mathrm{Mpc})^3$ run of the IllustrisTNG project, of which TNG-Cluster, with its  352 zoom-in clusters from a $(1~\mathrm{Gpc})^3$ box, is a spin off. Both TNG300+TNG-Cluster and FLAMINGO reproduce observed global X-ray properties of galaxy groups and clusters \citep{braspenning_flamingo_2024, nelson_introducing_2024}.

We use the sample of haloes from these two simulations to study the correlation between scatter in the X-ray luminosity ($L_{\rm X}$) -- halo mass ($M_{\rm 500c}$) scaling relation, and to elucidate the origin and evolution of the scatter, and to understand the physics driving it. Section~\ref{sec:methods} describes the simulations in detail and explains how scatter and correlation are defined. In Section~\ref{sec:results}, we quantify the scatter as a function of mass and redshift. Section~\ref{sec:mass-dependent-scatter} compares the scatter in FLAMINGO and TNG300+TNG-Cluster and studies the mass dependence. In Section~\ref{sec:redshift-dependent-scatter} we fit broken power laws to the redshift evolution of the scatter and quantify their accuracy.
Section~\ref{sec:results_2} presents the second part of the results where we try to understand the origin of the scatter and quantify the correlation between the scatter in different scaling relations.
Sections \ref{sec:correlate_with_second} and \ref{sec:reduce_scatter} use the $(2.8~\mathrm{Gpc})^3$ volume to reduce the scatter and provide fitting functions, Section~\ref{sec:gas_fraction} shows the effect of baryonic feedback variations, Section~\ref{sec:redshift} the evolution with redshift, and Section~\ref{sec:tng_comparison} the comparison with the TNG300+TNG-Cluster simulation suite. Finally, we provide our conclusions in Section~\ref{sec:conclusions}.

\section{Methods}\label{sec:methods} 
This section introduces the FLAMINGO and TNG300+TNG-Cluster simulations in Section~\ref{sec:methods_simulations}, then defines some of the properties used in this work in Section~\ref{sec:methods_properties}. The creation of median scaling relations and the definition of deviations therefrom are described in Section~\ref{sec:methods_correlation}. Finally, Section~\ref{methods:partial_correlation} describes the partial correlation analysis we use to explain the scatter of the observable-mass relation.

\subsection{Simulations} \label{sec:methods_simulations}
In this section we describe the cluster simulations, the relevant subgrid models, and the choice of cosmology and structure finding method.
\subsubsection{FLAMINGO}
FLAMINGO (Full-hydro Large-scale structure simulations with All-sky Mapping for the Interpretation of Next Generation Observations) is a large suite of hydrodynamical cosmological simulations that includes variations in baryonic feedback and cosmology, as well as three different resolutions and box sizes. One of the flagship runs has a volume of $\mathrm{(2.8 ~Gpc)^3}$ with a gas particle mass of $m_{\rm gas} = 1.07 \times 10^9 ~\mathrm{M_\odot}$, consisting of $2\times 5040^3$ gas and dark matter particles and $2800^3$ neutrino particles. This combination of volume and resolution makes it ideal for the statistical studies of populations of galaxy groups and clusters. The simulations are described in detail in \citet{schaye_flamingo_2023}. The machine learning-aided calibration of the stellar and AGN subgrid feedback parameters to the $z=0$ galaxy stellar mass function and $z=0.1-0.3$ cluster gas fractions \citep{kugel_flamingo_2023} is a unique feature. Variations on the fiducial model shift the observed gas fractions up and down by a multiple of their uncertainty ($\sigma$) and recalibrate the model to fit those new data points. FLAMINGO also includes models that vary the galaxy mass function or the cosmology. We will not use the latter here as we found the cosmology variations do not affect the results (see Appendix \ref{sec:cosmology}).

In this work, we use the $z=0$ to $z=2$ snapshots of the  $\mathrm{(2.8 ~\mathrm{Gpc})^3}$ and $\mathrm{(1 ~\mathrm{Gpc})^3}$ simulations with the fiducial feedback model (hereafter, L2p8 and L1\_m9, respectively), which both have an initial gas particle mass $m_{\rm gas} = 1.07 \times 10^9 ~\mathrm{M_\odot}$ and a dark matter particle mass $m_{\rm CDM} = 5.65 \times 10^9 ~\mathrm{M_\odot}$. Furthermore, we use the gas-fraction variations $f_{\rm gas}+2\sigma$ and $f_{\rm gas}-8\sigma$ in a $\mathrm{(1 ~\mathrm{Gpc})^3}$ volume with the same resolution \citep[see][for details on the gas fraction calibration]{kugel_flamingo_2023}.

FLAMINGO uses the open source simulation code \textsc{swift} \citep{schaller_swift_2018} and solves the hydrodynamics using the \textsc{sphenix sph} scheme \citep{borrow_sphenix_2022}, includes massive neutrions with the $\delta$f method \citep{elbers_optimal_2021}, and has initial conditions generated with a modified version of \textsc{monofonic} \citep{hahn_higher_2021, elbers_higher_2022}, assuming the cosmology `3x2pt + all external constraints' from the Dark Energy Survey year three analysis: $\Omega_{\rm m} = 0.306$, $\Omega_{\rm b} = 0.0486$, $\sigma_8 = 0.807$, $\mathrm{H_0} = 68.1~\mathrm{km~s^{-1}~Mpc^{-1}}$, $n_{\rm s} = 0.967$ \citep{abbott_dark_2022}.

The FLAMINGO model includes subgrid implementations of radiative cooling \citep{ploeckinger_radiative_2020}, star formation \citep{schaye_relation_2008}, stellar mass loss \citep{wiersma_chemical_2009, schaye_eagle_2015}, supernova feedback \citep{chaikin_importance_2022}, seeding and growth of SMBHs, and thermally-driven AGN feedback \citep{springel_modelling_2005, booth_cosmological_2009, bahe_importance_2022}. Additionally, two variations use kinetic jet feedback from AGN \citep{husko_spin-driven_2022}.

Cosmological structure identification is done using a modified version of the Hierarchical Bound Tracing algorithm \citep[HBT-HERONS;][]{forouharmoreno_assessing_2025, han_hbt_2018}, leveraging hierarchical structure formation by tracing objects through time, which leads to more robust substructure identification than for traditional halo finders. After running a 3D Friends-of-Friends (FoF) algorithm on the dark matter particles, HBT-HERONS finds haloes using an iterative unbinding procedure, and particles attached to self-bound objects are tracked across time. We further process the HBT-HERONS catalogues using the Spherical Overdensity and Aperture Processor \citep[SOAP;][]{mcgibbon_soap_2025} to compute halo properties within a range of apertures centred on the most bound particle (hereafter called the center-of-potential). For this work, we use two kinds of properties. First, the properties based on all particles within the spherical overdensity radius $r_{\rm 500c}$, defined as the radius within which the average enclosed density is 500 times the critical density of the universe. Second, we use bound subhalo properties, which only include particles bound to a specific subhalo. We only consider central subhaloes, which observationally are often brightest cluster galaxies (BCGs).

\begin{table} 
    \centering
    \caption{Halo properties ($\mathcal{Y}$) considered in this study. The first column gives the property symbol, the second column a short description, and the third column the units. The first section are ICM properties, the second section galaxy properties, and the third section are halo properties that can also be computed for dark matter only simulations.}
    \begin{tabular}{l|l|l} \label{tab:all_properties}
        $\mathcal{Y}$ & description & units\\ \hline
        
        $f_{\rm gas}$ & Gas mass fraction & - \\
        Y & Compton-Y parameter & $\mathrm{{Mpc}^{-2}}$ \\
        $T_{\rm gas}$ & Mass-weighted gas temperature & $\mathrm{K}$ \\
        $\sigma_{\rm gas}$ & Gas velocity dispersion & $\mathrm{km ~ s^{-1}}$ \\
        Z & Total metal mass fraction in gas & -\\ 
        $E_{\rm kin} / E_{\rm therm}$ & Kinetic-to-thermal energy ratio for gas & - \\
        \hline
        $f_\star$ & Stellar mass fraction & - \\
        SFR & Star formation rate & $\mathrm{M_\odot ~ {yr}^{-1}}$ \\
        $M_{\rm MMBH}$ & Most massive black hole mass & $\mathrm{M_\odot}$ \\
        $\dot{M}_{\rm MMBH}$ & Most massive black hole accretion rate & $\mathrm{M_\odot ~ {yr}^{-1}}$ \\
        $\sigma_\star$ & Stellar velocity dispersion & $\mathrm{km ~ s^{-1}}$ \\
        $t_\star$ & Mean mass-weighted stellar age & $\mathrm{Gyr}$ \\
        \hline
        $f_{\rm sat}$ & Mass fraction in satellites & - \\
        $c_{\rm DM}$ & Dark matter concentration & - \\
        $\Delta_{\rm x}$ & COM-COP offset & $r_{\rm 500c}$ \\
        $\sigma_{\rm DM}$ & Dark matter velocity dispersion & $\mathrm{km ~ s^{-1}}$ \\
        $\lambda$ & Spin & - \\
        $s$ & Sphericity & - \\
        $t$ & Triaxiality & - \\
        $e$ & Ellipticity & - \\
        $M_{\rm 500c} / M_{\rm 200c}$ & Ratio of masses within $r_{\rm 500c}$ and $r_{\rm 200c}$ & - \\
        $M_{\rm 2500c} / M_{\rm 500c}$ & Ratio of masses within $r_{\rm 2500c}$ and $r_{\rm 500c}$ & - \\
        $v_{\rm max}$ & Maximum circular velocity & $\mathrm{km ~ s^{-1}}$\\ \hline  
    \end{tabular}
\end{table}

\subsubsection{TNG-Cluster and TNG300}
TNG-Cluster is described in detail in \citet{nelson_introducing_2024} and is a suite of 352 zoom simulations of galaxy clusters selected from a $(1~\mathrm{Gpc})^3$ parent volume. It was designed to supplement the high-mass end of TNG300. The minimum mass of TNG-Cluster selected haloes is $M_{\rm 200c} = 10^{14.3} ~ \mathrm{M_\odot}$, with haloes selected at random in $0.1~\mathrm{dex}$ mass bin up to $M_{\rm 200c} = 10^{15.0} ~ \mathrm{M_\odot}$ to compensate for the drop-off in statistics from TNG300. This results in a flat mass function up to $M_{\rm 200c} = 10^{15.0} ~ \mathrm{M_\odot}$, whereas above this all haloes in the parent box were selected for re-simulation, meaning the sample is volume limited \citep{nelson_introducing_2024}. 

Both TNG300 and TNG-Cluster use the IllustrisTNG galaxy-formation model \citep{weinberger_simulating_2017, pillepich_simulating_2018}, with a target baryonic mass resolution of $1.2 \times 10^7 ~\mathrm{M_\odot}$ and a dark matter particle mass of $m_{\rm CDM}=6.1 \times 10^7 ~\mathrm{M_\odot}$. Structure is identified using the \textsc{subfind} algorithm \citep{springel_populating_2001, dolag_substructures_2009}, which finds gravitationally bound sets of particles and cells at a given cosmic time within previously identified Friends-of-Friends haloes.

In this work, we use the $z=0$ to $z=2$ snapshots of all 352 haloes in TNG-Cluster and the 2548 haloes (at $z=0$) in TNG300 with $M_{\rm 500c} > 10^{13}~\mathrm{M_\odot}$, for a total sample (at $z=0$) of 2900 haloes. Similar to FLAMINGO, we use both properties within a spherical radius $r_{\rm 500c}$ and those from stellar resolution elements that are bound to the central subhalo. Unlike FLAMINGO, the center of a halo is defined as the particle with the lowest gravitational potential energy (hereafter called the center-of-potential). The IllustrisTNG model uses the cosmology from \citet{planck_collaboration_planck_2016}: $\Omega_{\rm m} = 0.3089$, $\Omega_{\rm b} = 0.0486$, $\sigma_8 = 0.8159$, $\mathrm{H_0} = 67.74~\mathrm{km~s^{-1}~Mpc^{-1}}$, $n_{\rm s} = 0.9667$).

Comparing the outcome of FLAMINGO with that of the IllustrisTNG galaxy-formation model offers an independent test of our results with a simulation that uses a different hydrodynamic solver and different physical models for galaxy-physics processes, from cooling, to stellar and AGN feedback. 

\begin{figure*}
    \centering
    \includegraphics[width = \linewidth]{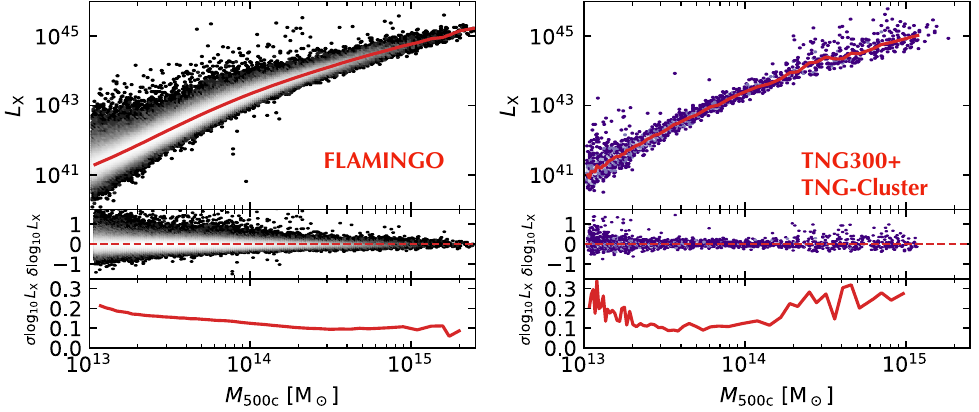}
    \caption{The X-ray luminosity -- halo mass ($L_{\rm X}$ -- $M_{\rm 500c}$) relation over two decades of halo mass for all galaxy groups and clusters in the L2p8 FLAMINGO simulation (left) and TNG300 + TNG-Cluster (right) at $z=0$ (top panels). Red curves give the medians in narrow mass bins. The logarithmic deviation from the median scaling relation (in dex) is shown in the middle panel (Eq.~\ref{eq:log_deviation}). The bottom panels show the scatter in the relation (in dex), namely the quantity we focus on throughout this paper and defined as per Eq.~\ref{eq:log_scatter}. Whereas the scatter decreases monotonically with mass for FLAMINGO, from $0.2$ to $0.1$ dex, it increases above $10^{14}~\mathrm{M_\odot}$ for the TNG300+TNG-Cluster simulations, ranging from 0.1 to 0.3 dex.}
    \label{fig:scaling_relation}
\end{figure*}

\subsection{Definition of halo properties} \label{sec:methods_properties}
For this work, we use all central haloes with mass $M_{\rm 500c} > 10^{13}~\mathrm{M_\odot}$ (of which there are 2,076,175 in L2p8, 94,374 in L1\_m9, and 2900 in TNG300+TNG-Cluster): no relaxation criterion is applied, but we require haloes to have at least 100 bound gas volume elements. 

We measure gas properties in 3D spheres with radius $r_{\rm 500c}$. We list all the properties we consider throughout this paper in Table~\ref{tab:all_properties}, where we distinguish between ICM properties in the first section, non-gas galaxy properties in the second section, and properties that can also be measured from a dark matter only simulation in the bottom section. We now define the non-trivial properties.

The X-ray luminosity in this work is the intrinsic $0.5-2.0~\mathrm{keV}$ broad band luminosity [$\mathrm{erg}~\mathrm{s^{-1}}$] in the observer frame, without any instrumental response function applied. The luminosity is calculated for each volume element based on interpolating output from the photo-ionisation spectral synthesis code Cloudy \citep{ferland_2017_2017} in redshift, density, temperature, and 9 individual element abundances. A detailed description is given in \citet{braspenning_flamingo_2024}. 

The thermal Sunyaev-Zel'dovich (SZ) Compton-Y is computed by summing over the Compton-Y contributions from individual volume elements,
\begin{equation}
    y = \frac{\sigma_{\rm T}}{m_{\rm e} c^2}n_{\rm e} k_{\rm B} T_{\rm e} \frac{m}{\rho} \, ,
\end{equation}
with $\sigma_{\rm T}$ being the Thomson cross section, $m_{\rm e}$ the electron mass, $c$ the speed of light, $n_{\rm e}$ the electron number density, $T_{\rm e}$ the electron temperature, $m$ the mass and $\rho$ the density of a particle. The electron number density is computed using the cooling table (\citealt{ploeckinger_radiative_2020} for FLAMINGO; \citealt{wiersma_effect_2009} for IllustrisTNG).

Thermal AGN feedback in FLAMINGO heats a single particle to a high temperature. Such particles are typically close to the cluster centre and have high densities. In the short time before the particle can adiabatically expand and transfer its energy to nearby particles, the high density combined with high temperature can dominate the total X-ray emission of a halo, which would be unphysical. Therefore, we exclude gas that has experienced direct AGN heating in the last $15 ~{\rm Myr}$ for the X-ray luminosity and Compton-Y signal. We have confirmed that this makes no significant difference for the results in the work. No such choice is applied (or could be applied) to TNG300+TNG-Cluster.

For the velocity dispersion of each particle species, we use the dispersion along a single-axis
\begin{equation}
    \sigma^2 = \frac{1}{\sum_{i=1}^n m_i} \sum_{i=1}^n m_i v_{x, i} v_{x, i} \, ,
\end{equation}
with $v_{x, i}$ the velocity of a particle along one axis relative to the centre of mass velocity of the halo. We separately use the velocity dispersion of gas $\sigma_{\rm gas}$, stars $\sigma_\star$, and dark matter $\sigma_{\rm DM}$.

The spin parameter is computed following \citet{bullock_universal_2001}
\begin{equation}
    \lambda = \frac{|\Vec{L}_{\rm tot}|}{\sqrt{2} M_{\rm 500c} v_{\rm max} r_{\rm 500c}} \, 
\end{equation}
in which $\Vec{L}_{\rm tot}$ is the total angular momentum of all particles within radius $r_{\rm 500c}$, and $M_{\rm 500c}$ their total mass, while $v_{\rm max}$ is the maximum circular velocity within the same radius. The angular momentum is computed relative to the most bound particle of the halo and the halo's center-of-mass velocity.

The sphericity, ellipticity, and triaxiality are calculated from the reduced inertia tensor computed in a single iteration from the total mass distribution. Denoting the three eigenvalues of the reduced inertia tensor as ($\lambda_0, \lambda_1, \lambda_2$), the shape properties are defined as
\begin{align}
    \text{sphericity} &= s = \frac{\lambda_0}{\lambda_2} \\
    \text{ellipticity} &= e = \frac{\lambda_1}{\lambda_2} \\
    \text{triaxiality} &= t = \frac{{\lambda_0}^2 - {\lambda_1}^2}{{\lambda_0}^2 - {\lambda_2}^2}
\end{align}

\begin{figure*}
    \centering
    \includegraphics[width=  \linewidth]{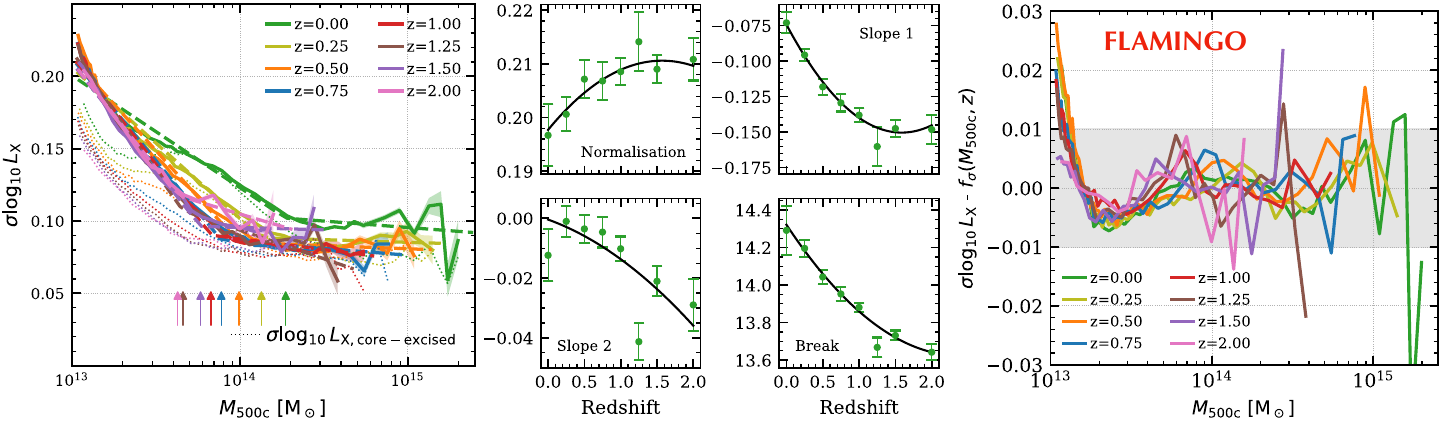}
    \includegraphics[width=  \linewidth]{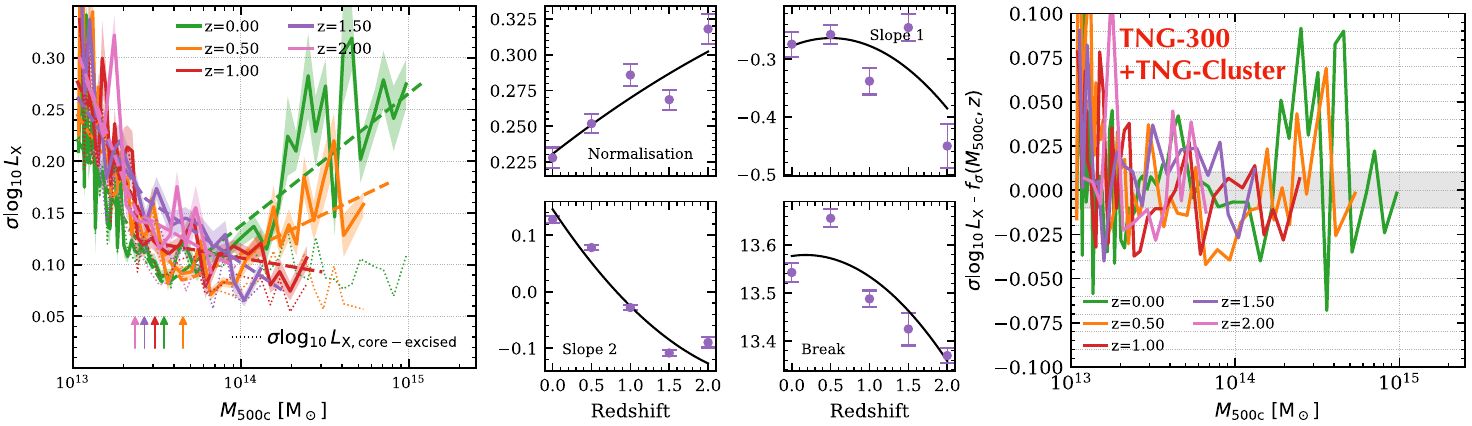}
    \caption{Redshift evolution of the scatter in the $L_{\rm X}$ -- $M_{\rm 500c}$ relation for FLAMINGO $\mathrm{L2p8}$ (top) and TNG300+TNG-Cluster (bottom). The left panels show the scatter as a function of mass from $z=0$ to $z=2$ (solid curves), with shaded areas depicting the $1\sigma$ bootstrap error in each bin. The dotted curves show the scatter for the core-excised luminosity $L_{\rm X, ce}$, which only differs significantly from the scatter in the total luminosity either below $10^{14}~\mathrm{M_{\odot}}$ (FLAMINGO) or above that mass (TNG300+TNG-Cluster). A best-fit broken power law for each redshift is shown as dashed lines of the same colour, with arrows indicating the break point. The middle four-tile panels show the redshift evolution of the four parameters of the broken power law (Eq.~\ref{eq:broken_powerlaw_scatter}), and the best fit quadratic function (Eq.~\ref{eq:redshift_evolution_broken_powerlaw}) to that evolution. The right most panels show the difference between the true scatter in each bin, and the prediction from the broken power law using the parameters at each redshift from the best quadratic fit shown in the middle panels. The quadratic fit to the redshift evolution, combined with the broken power law, allows an accurate prediction, within $0.01~\mathrm{dex}$ (FLAMINGO) and $0.05~\mathrm{dex}$ (TNG300+TNG-Cluster), across a wide range of redshift and mass.}
    \label{fig:scatter_redshift_FLAMINGO_TNG}
\end{figure*}

\subsection{The $L_{\rm X}$ -- $M_{\rm 500c}$ relation and definition of its scatter} \label{sec:methods_correlation}
We first define the scaling relation in X-ray luminosity. We create the median scaling relation by sorting all haloes by mass and then dividing them into bins of 30,000 objects (for L2p8), 1500 objects (for L1\_m9) or 50 objects (for TNG300+TNG-Cluster). At the high-mass end, we iteratively refine the bins until the objects in a single bin span less than $0.1~\mathrm{dex}$ in halo mass, or the lower limit of 20 objects in a bin is reached (for L2p8 the lower object limit is never reached at $z=0$). This results in $\approx$$50$ cluster-mass bins. Instead of fitting a power-law scaling relation, we calculate the median of the X-ray luminosity $L_{\rm X}$ in each halo mass bin. 

Using this median, we then compute the logarithmic deviation for each object
\begin{equation} \label{eq:log_deviation}
    \delta \log_{10} L_{\rm X} = \log_{10} L_{\rm X,i} - \overline{\log_{10} L_{\rm X}} \, ,
\end{equation}
with $\overline{\log_{10} L_{\rm X}}$ the median $\log_{10} L_{\rm X}$-value of the scaling relation at the same mass $M_{\rm 500c}$ as halo $i$. Hence $\overline{\log_{10} L_{\rm X}} = \text{median}(\log_{10}L_{\rm X}(M_{\rm 500c}))$, where the median is found with a cubic spline interpolation between medians computed in bins, such that each halo uses the median at its exact halo mass. $\delta \log_{10} L_{\rm X}$ is thus the logarithmic deviation (in $\mathrm{dex}$) from the median scaling relation. 

We define the scatter of a mass bin as
\begin{equation}\label{eq:log_scatter}
    \sigma \log_{10} L_{\rm X} = \frac{1}{2} \left(P_{84}(\delta \log_{10} L_{\rm X}) - P_{16}(\delta \log_{10} L_{\rm X}) \right) \, ,
\end{equation}
with $P_{x}$ the x-th percentile of the distribution within that bin.

For every halo, we compute its $\delta \log_{10} L_{\rm X}$, and for all properties $\mathcal{Y}$ listed in Table~\ref{tab:all_properties} the $\delta \log_{10} \mathcal{Y}$, allowing us to correlate deviations between different scaling relations for the same halo.

\subsection{Computing partial correlations} \label{methods:partial_correlation}
In this work we explain the scatter of $L_{\rm X}$ at fixed mass $M_{\rm 500c}$ by accounting for the dependence of $L_{\rm X}$ on the deviation of a second property $\mathcal{Y}$ from its scaling relation:
\begin{equation}
    L_{\rm X}(\delta \log_{10} \mathcal{Y} | M_{\rm 500c} = \text{const.}) \, .
\end{equation}
If there is a strong dependence of $L_{\rm X}$ on $\delta \log_{10} \mathcal{Y}$, then accounting for it could reduce the scatter in $L_{\rm X}$. 

In practice we fit a linear relation at fixed $M_{\rm 500c}$ between the deviations in $\mathcal{Y}$ and in $L_{\rm X}$
\begin{equation}
        \delta \log_{10} L_{\rm X} (\delta\log_{10}\mathcal{Y}) = \alpha + \beta \times \delta \log_{10}\mathcal{Y} \, , \label{eq:fit_second_property_log}
\end{equation}
with $\alpha$ and $\beta$ free parameters. These are fit using the same bins as defined before, namely at most 30,000 (L2p8), 1500 (L1\_m9), or 50 (TNG300+TNG-Cluster) objects per bin and smaller than $0.1 ~\mathrm{dex}$ in $M_{\rm 500c}$. The mass dependence of $\alpha$ and $\beta$ is subsequently fit using a third-order polynomial
\begin{equation} \label{eq:fit_second_property_log_mass} 
    \alpha(M_{\rm 500c}) \text{ or } \beta(M_{\rm 500c}) = c_1 + c_2 ~ M_{\rm 500c} + c_3 ~ M_{\rm 500c}^2 + c_4 ~ M_{\rm 500c}^3
\end{equation} 

For each halo, we calculate the residual scatter after removing the part that can be explained by invoking the quantity $\mathcal{Y}$,
\begin{align}
    \delta &\log_{10}L_{\rm X}' \\
    &= \delta \log_{10}L_{\rm X} - \delta \log_{10}L_{\rm X} (M_{\rm 500c}, \delta \log_{10}\mathcal{Y} )\\
    &= \delta \log_{10}L_{\rm X} - \left[ \alpha(M_{\rm 500c}) + \beta(M_{\rm 500c}) \times \delta\log_{10}\mathcal{Y} \right] \, . \label{eq:correction_second_property_log}
\end{align}
This analysis can be done for the combination of $L_{\rm X}$ with any of the properties $\mathcal{Y}$ from Table~\ref{tab:all_properties}.

\section{Scatter in the X-ray luminosity -- halo mass relation} \label{sec:results}
In this section we present results on the scatter in the X-ray luminosity -- halo mass relation for FLAMINGO and TNG300+TNG+Cluster. In Section~\ref{sec:mass-dependent-scatter} we first show the scaling relation for both simulations, and the logarithmic scatter around it as a function of cluster mass. We then focus on the redshift evolution of the scatter, and fit it with simple analytic functions in Section~\ref{sec:redshift-dependent-scatter}.

\subsection{Mass dependence of the scatter} \label{sec:mass-dependent-scatter}
The top two panels of Fig.~\ref{fig:scaling_relation} show the X-ray luminosity ($L_{\rm X}$) -- halo mass ($M_{\rm 500c}$) relation for all objects with $M_{\rm 500c} > 10^{13} ~ \mathrm{M_\odot}$ in the FLAMINGO $(2.8~\mathrm{Gpc})^3$ volume (left) and the combination of TNG300 and TNG-Cluster (right). A lighter colour indicates a larger number of haloes. The red line shows the median scaling relation, as defined in Section~\ref{sec:methods_correlation}. The middle panels show the logarithmic deviation (Eq.~\ref{eq:log_deviation}) from the scaling relations, computed using an interpolation between the binned median relation to the exact mass of each halo. The bottom panels show the mass-dependent scatter in dex, which we define as the difference between the $84^{\rm th}$ and $16^{\rm th}$ percentiles, computed from the middle panels (Eq.~\ref{eq:log_scatter}). 

In both simulations, more massive haloes have a higher luminosity, as expected. For FLAMINGO the scatter decreases with increasing halo mass. At $10^{13}~\mathrm{M_\odot}$ the scatter is $\approx 0.2~\mathrm{dex}$, decreasing to $\approx 0.1~\mathrm{dex}$ for $10^{14.25}~\mathrm{M_\odot}$, after which it remains almost constant. For TNG300+TNG-Cluster the scatter initially decreases towards more massive clusters, until $10^{14}~\mathrm{M_\odot}$, after which it rises again. This is already directly visible from the top panel, where the most massive haloes show a spread in luminosities that is larger than what is seen in FLAMINGO, even though the latter as a 27$\times$ larger volume.

\subsubsection{Notes on the scatter about a fitted power law}
Fitting a power law is a popular method for characterizing scaling relations. This is also often done for the X-ray luminosity -- halo mass scaling relation. While performing that fit, one can simultaneously add a parameter representing the population average scatter, resulting in the minimization of
\begin{equation}
    \chi^2 = \sum_i \frac{\left( \log_{10} L_{\mathrm{X}, i} - A - B \times \log_{10} M_{\mathrm{500c}, i} \right)^2}{\sigma^2} \, ,
\end{equation}
where $A$ and $B$ are the parameters of the power law, and $\sigma$ is the scatter. When fit to a volume-limited sample of objects, this is naturally biased toward lower mass haloes, as those are more abundant and hence contribute more to the minimization procedure. Hence, the scatter we obtain in this way over-estimates the true scatter for high-mass haloes.

Another problem with this approach is that a power law is a good approximation of the scaling relation only on average. The actual median $L_{\rm X}$ tends to be slightly below the scaling relation at low mass ($10^{13}~\mathrm{M_\odot}$) and slightly above at intermediate masses ($10^{14}~\mathrm{M_\odot}$), due to baryonic processes moving haloes away from the perfect power-law relations in dark matter-only universes. Such over- and under estimates of the true median result in artificial enhancements of the best-fitting scatter parameter compared to the true scatter.
We perform this exercise at $z=0$, and find that the scatter from fitting a power law across the full mass range is $\sigma_{L_{\rm X}} = 0.25$ and $\sigma_{L_{\rm X}} = 0.34$ for FLAMINGO and TNG300+TNG-Cluster respectively, larger even than the maximum scatter we find in Fig.~\ref{fig:scaling_relation}. 

Given these limitations, we do not pursue this approach any further. We point out that the mass-dependent scatter of this work is agnostic to the definition of the scaling relation as it measures the spread in each bin, independent of the choice of centre. However, defining the spread in each bin as the width of a lognormal does systematically overestimate the scatter by 5-10\% (see Appendix \ref{sec:appendix_lognormal}) in line with findings from \citet{kugel_flamingo_2024} that a lognormal only approximately describes the distribution of the scatter in $L_{\rm X}$ -- $M_{\rm 500c}$.

\subsection{Redshift evolution of the scatter} \label{sec:redshift-dependent-scatter}
Having established the scaling relation and scatter at $z=0$, we study the evolution of the scatter with redshift. For this we do a side-by-side comparison of the FLAMINGO and IllustrisTNG models. For the former we use the largest volume ($2.8~\mathrm{Gpc}$)$^3$, and for the latter the combination of haloes from TNG300 and TNG-Cluster, selecting in both cases all haloes with $M_{\rm 500c} \geq 10^{13}~\mathrm{M_\odot}$. 

We fit the scatter as a function of halo mass, in both simulations and at each redshift, with a broken power law
\begin{equation}
    f_\sigma (M_{\rm 500c}, z) = 
    \begin{cases}
        A(z) + \alpha_1(z) ~ \log_{10} M_{\rm 500c}^\star & M_{\rm 500c}^\star < M_{\rm break}^\star(z) \\
        A'(z) + \alpha_2(z) ~ \log_{10} M_{\rm 500c}^\star & M_{\rm 500c}^\star \geq M_{\rm break}^\star(z)
    \end{cases} \, ,
    \label{eq:broken_powerlaw_scatter}
\end{equation}
where $M_{\rm 500c}^\star = M_{\rm 500c} / 10^{13}~\mathrm{M_\odot}$, the normalisation of the second power law, is fixed by requiring continuity 
\begin{equation}
    A'(z) = A(z) + \log_{10} M_{\rm break}^\star(z) \left[\alpha_1(z) - \alpha_2(z)\right] \, .
\end{equation}
We model the redshift evolution of each power law parameter with simple quadratic function. For example, the normalisation
\begin{equation}
    A(z) = \beta_1 ~ (1+z)^2 + \beta_2 ~ (1+z) + \beta_3 \, .
    \label{eq:redshift_evolution_broken_powerlaw}
\end{equation}

\begin{table}
    \centering
    \caption{Best-fit parameters for the redshift evolution (Eq.~\ref{eq:redshift_evolution_broken_powerlaw}) of the broken power law (Eq.~\ref{eq:broken_powerlaw_scatter}) which can be used to describe the scatter of the $L_{\rm X}$ -- $M_{\rm 500c}$ relation according to the FLAMINGO (top) and TNG300+TNG-Cluster (bottom) simulations.}
    \begin{tabular}{l|l|l|l} \hline
         & $\beta_1$ & $\beta_2$ & $\beta_3$ \\ \hline
       \textbf{FLAMINGO} & & & \\ \hline
       normalization ($A$) & -0.0052 & 0.0269 & 0.1761\\
       slope 1 ($\alpha_1$) &  0.0302 & -0.1563 & -0.0518\\
       slope 2 ($\alpha_2$) & -0.0045 & -0.0003 & -0.0038\\
       break-point ($M_{\rm break}$) & 0.1222 & -0.8281 & 15.0273 \\ \hline
       \textbf{TNG300+TNG-Cluster} & & & \\ \hline
       normalization ($A$) & -0.0034 & 0.0491 & 0.1849\\
       slope 1 ($\alpha_1$) &  -0.0546 & 0.1642 & -0.3867\\
       slope 2 ($\alpha_2$) & 0.0350 & -0.2768 & 0.3878\\
       break-point ($M_{\rm break}$) & -0.0665 & 0.1567 & 13.4868\\ \hline
    \end{tabular}
    \label{tab:best_fit_scatter_redshift_params}
\end{table}

\begin{figure*}
    \centering
    \includegraphics[width=\linewidth]{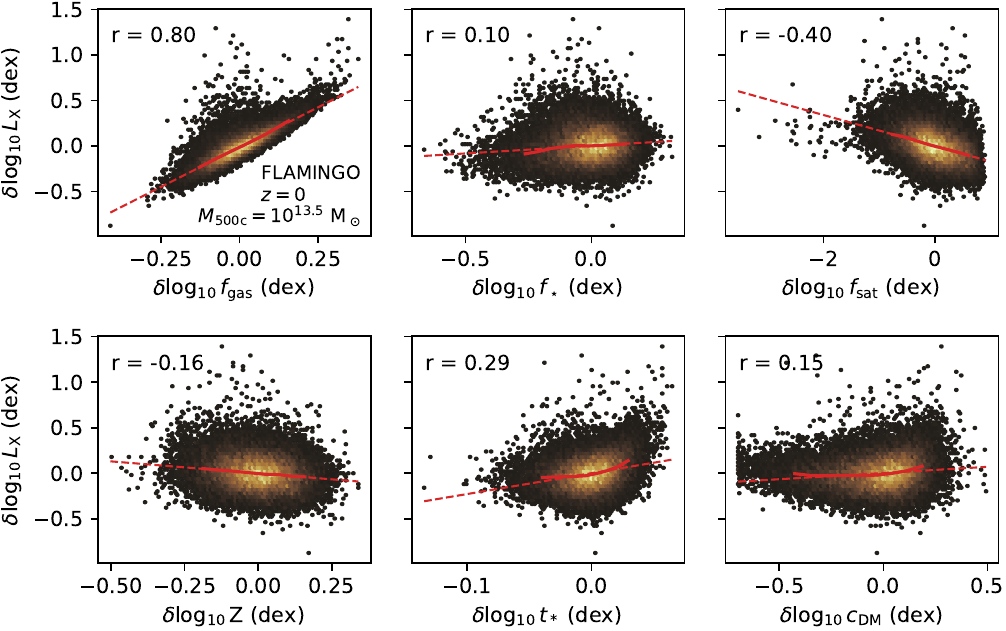}
    \caption{The logarithmic deviation from the $L_{\rm X}$ -- $M_{\rm 500c}$ scaling relation (in dex; Eq.~\ref{eq:log_deviation}) against the logarithmic deviation in a second cluster property (selected from Table~\ref{tab:all_properties}) for FLAMINGO L2p8 at $z=0$ at $M_{\rm 500c} = 10^{13.5} ~ \mathrm{M_\odot}$ (30,000 haloes). Lighter colours indicate a larger number of haloes. The red error bars (barely visible) are 16th-84th percentile bootstrap resamplings of the median in ten logarithmic bins of the second property. The dashed red line is the best linear fit for those error bars. The spearman rank correlation is indicated in the top left of each panel. Some of the properties show a clear correlation with the deviation from the X-ray luminosity scaling relation, but the level of correlation and the scatter differ strongly between properties.}
    \label{fig:second_property_137}
\end{figure*}

\begin{figure*}
    \centering
    \includegraphics[width=\linewidth]{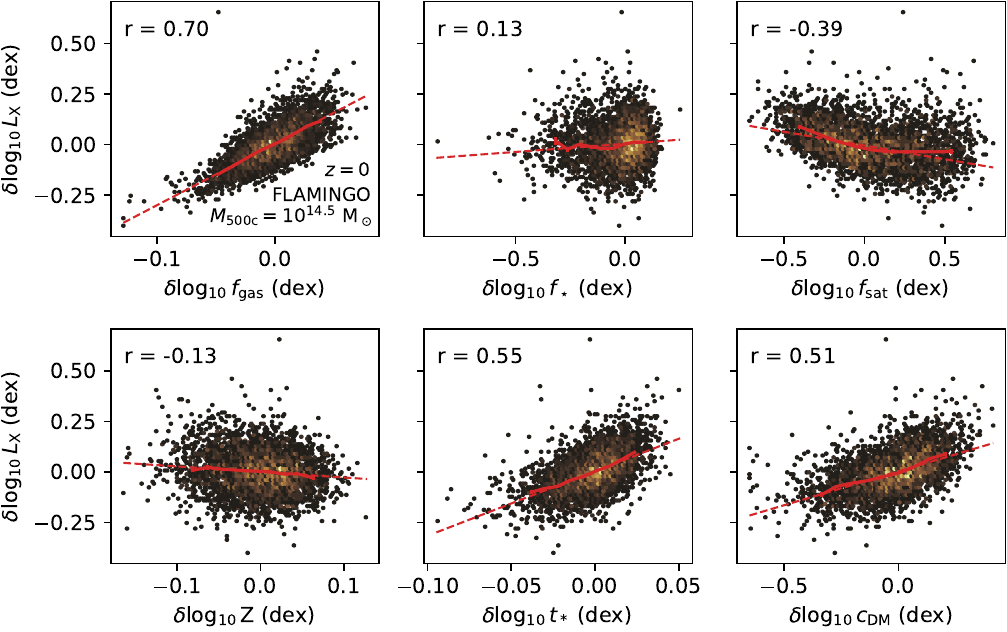}
    \caption{As Fig.~\ref{fig:second_property_137}, for FLAMINGO L2p8 at $z=0$, but for the higher-mass clusters with $M_{\rm 500c} = 10^{14.5} ~ \mathrm{M_\odot}$ (3771 haloes).}
    \label{fig:second_property_147}
\end{figure*}

The two left panels of Fig~\ref{fig:scatter_redshift_FLAMINGO_TNG} show the scatter in $L_{\rm X}$--$M_{\rm 500c}$ for FLAMINGO (top) and TNG300+TNG-Cluster (bottom) between $z=0$ and $2$. Both simulations show redshift evolution, but in a markedly different manner. For FLAMINGO, the low-mass slope of the scatter gradually steepens with increasing redshift, and after the break-point, which shifts to lower mass with increasing redshift, the scatter is almost constant with halo mass. On the other hand, haloes from the TNG300+TNG-Cluster simulations show a stark increase in the scatter at $z=0$ above $M_{\rm 500c} \approx 10^{14}~\mathrm{M_\odot}$, which becomes less prominent towards $z=1$, above which redshift this mass range is no longer probed by the TNG300+TNG-Cluster simulations. The low-mass slope, at every redshift, is much steeper according to the IllustrisTNG model than for FLAMINGO, and the break-point only shows a very weak redshift evolution. 

The middle four panels of each row show the parameters of the best-fit broken power law at each redshift as error bars. For both simulations, the redshift evolution is well fit by a simple quadratic function, shown as a solid black line (parameters of the best fits are in Table~\ref{tab:best_fit_scatter_redshift_params}).

The right-most panel in each row shows the difference between the true scatter of each bin, and the predicted scatter from the broken power law using at each redshift the parameters from the quadratic fits shown in the middle panels (i.e. not the best fitting values at each individual redshift as shown by the error bars). For FLAMINGO, the fit recovers the scatter within $0.01~\mathrm{dex}$ over 2 decades in mass and from $z=0$ to $z=2$. The same result for TNG300+TNG-Cluster is harder to quantify due to the large fluctuations in the scatter towards high masses  and lower redshifts. However, the error in the scatter rarely exceeds $0.05~\mathrm{dex}$ . The large bin-to-bin fluctuations in TNG300+TNG-Cluster can be explained by the fact that there are 600$\times$ fewer haloes per mass bin compared to FLAMINGO.

In the left-most panels of Fig~\ref{fig:scatter_redshift_FLAMINGO_TNG}, thin dotted lines show the scatter of the core-excised ($r > 0.15 r_{\rm 500c}$) X-ray luminosity, a quantity often used in observations. Our results show that the reduction in scatter due to core-excision is minimal over most of the mass range, with only two exceptions. For FLAMINGO we see a reduction in the scatter below $M_{\rm 500c}=10^{14}~\mathrm{M_\odot}$, with a maximum reduction of $0.05~\mathrm{dex}$ at $M_{\rm 500c}=10^{13}~\mathrm{M_\odot}$. For TNG300+TNG-Cluster, there is no difference in the scatter at these lower masses, but core-excision leads to a flat scatter--halo mass relation above $M_{\rm 500c}=10^{14}~\mathrm{M_\odot}$. The core-excision removes the growth in scatter at this massive cluster scale, hinting at small scale processes in the TNG300+TNG-Cluster simulations giving rise to large halo-to-halo variations. Visual inspection of 2D X-ray maps of the cores of a few selected objects reveals the presence of extremely X-ray bright small clumps in the central regions. Such clumps are not present in FLAMINGO, even when recently heated gas particles are not removed.

Despite the fact that the best fitting parameters are quite different between FLAMINGO and TNG300+TNG-Cluster, we notice qualitatively similar trends with redshift. 
The low-mass slope steepens with redshift such that the magnitude of the scatter at the break point is constant across redshift. We see that the second slope (above the mass of the break-point) also becomes more negative with redshift, indicating that there might be a mechanism at play elevating the scatter of the most massive objects at lower redshift. This is particularly clear for TNG300+TNG-Cluster, where the scatter increases, rather than decreases for the most massive clusters at $z=0$.

\section{Understanding the scatter in the X-ray luminosity -- halo mass relation} \label{sec:results_2}
\begin{figure}
    \centering
    \includegraphics[width=\linewidth]{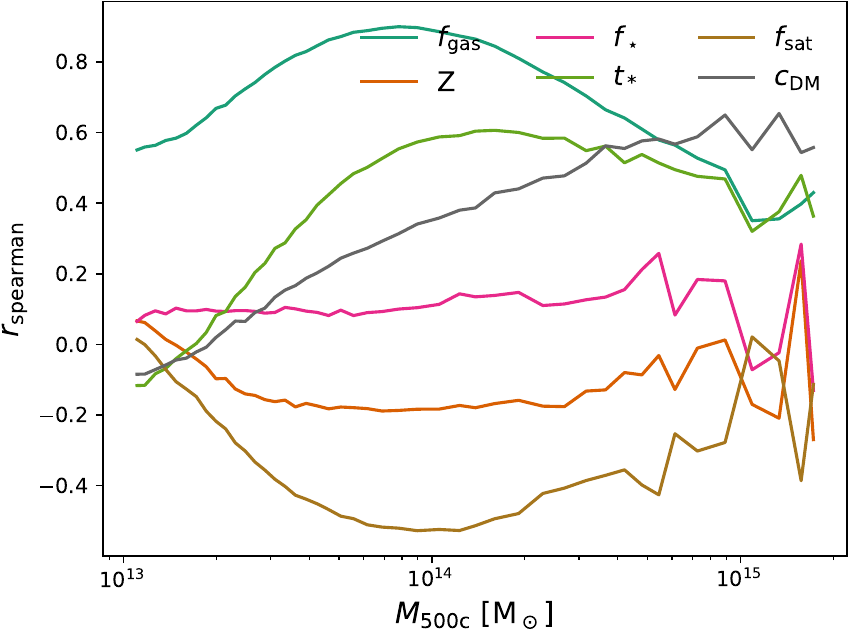}
    \caption{Mass-dependent Spearman-rank correlation coefficients between the logarithmic deviation from the $L_{\rm X}$ -- $M_{\rm 500c}$ scaling relation and the six cluster properties showcased in Figures~\ref{fig:second_property_137} and \ref{fig:second_property_147}.
    According to FLAMINGO, most quantities have a characteristic mass at which the correlation strength peaks. For $f_{\rm gas}$ this is around $10^{13.75}~\mathrm{M_\odot}$, and for  $c_{\rm DM}$ around $10^{15}~\mathrm{M_\odot}$, whereas for all other properties it is around $10^{14}~\mathrm{M_\odot}$, indicating that different mechanisms drive deviations in these quantities.}
    \label{fig:betas_mass_dependence}
\end{figure}

We now turn our attention to finding other halo properties, beyond X-ray luminosity and mass, which can yield a reduction of the scatter in the $L_{\rm X}$ -- $M_{\rm 500c}$ scaling relation. In this section we mostly focus on the FLAMINGO simulation, due to its much larger number of haloes, making our results statistically robust.

In Section~\ref{sec:correlate_with_second} we fit the deviation of each halo from different scaling relations, and provide all the best fitting parameter values. In Section~\ref{sec:reduce_scatter} we use these fits to show how much we can reduce the scatter in the X-ray luminosity scaling relation with knowledge of each additional property. We then show how this reduction depends on the FLAMINGO feedback variations in Section~\ref{sec:gas_fraction}, and on redshift in Section~\ref{sec:redshift}. We then apply our method to the TNG300+TNG-Cluster simulations and compare in Section~\ref{sec:tng_comparison}.

\subsection{Correlating the scatter with a second property} \label{sec:correlate_with_second}
Here, we correlate the scatter in the X-ray luminosity -- mass relation with a second halo property from Table~\ref{tab:all_properties}. We fit a linear function (Eq.~\ref{eq:fit_second_property_log}), in narrow halo mass bins, between the logarithmic deviation from the $L_{\rm X}$--$M_{\rm 500c}$ scaling relation ($\delta L_{\rm X}$) and the logarithmic deviation of a second property, $\delta \mathcal{Y}$, from its scaling relation $\mathcal{Y}$--$M_{\rm 500c}$. Using the deviation from the scaling relation of the second property, instead of the value of the second property itself, should erase any residual mass dependence in our already narrow bins.

Fig.~\ref{fig:second_property_137} shows the distribution of scatter against six secondary properties at $M_{\rm 500c} = 10^{13.5}~\mathrm{M_\odot}$, for all 30,000 haloes in a narrow mass bin (here $0.04~\mathrm{dex}$). The six properties are selected such that each category -- ICM, galaxy and dark matter -- is represented by two properties. It shows a strong positive correlation with the deviation from the gas fraction scaling relation, with a Pearson-rank correlation coefficient of $r = 0.8$. Namely, haloes with a higher gas fraction have a higher X-ray luminosity. Weaker correlations exist with, in order of decreasing correlation, satellite mass fraction ($r = -0.4$), BCG stellar age ($r = 0.29$), gas metallicity ($r = -0.16$), halo concentration ($r = 0.15$), and stellar mass fraction ($r = 0.1$). 

The strong correlation with gas fraction can easily be explained, as a more gas-rich halo will simply have more X-ray emitting material, resulting in a higher X-ray luminosity. The anti-correlation with satellite mass fraction likely relates to relaxedness. More relaxed haloes tend to have lower satellite mass fractions, higher concentrations, and be older \citep[e.g.][]{jeeson-daniel_correlation_2011}.

Fig.~\ref{fig:second_property_147} shows the same properties but for the mass bin at $10^{14.5}~\mathrm{M_\odot}$, containing 3771 haloes with a total bin width of $0.06~\mathrm{dex}$. The gas fraction still correlates strongly with X-ray luminosity ($r = 0.7$) at fixed halo mass, but now the dark-matter halo concentration ($r = 0.51$), and BCG stellar age ($r = 0.55$) also show a strong positive correlation, though there are large object-to-object variations. The correlation with stellar mass fraction ($r = 0.13$), gas metallicity ($r = -0.13$), and satellite mass fraction ($r = -0.39$) remain largely unchanged.

In each mass bin, we fit linear functions between $\delta \log_{10} L_{\rm X}$ and each $\delta \mathcal{Y}$. To increase robustness against outliers, we create ten logarithmic bins ranging from the 2nd to the 98th percentile of the second properties range (x-axis) within each mass bin. We calculate the median and an error on the median in each such bin using 10,000 bootstrap samples. Red error bars in Figures~\ref{fig:second_property_137} and \ref{fig:second_property_147} indicate the medians and the error on the medians. The red dashed line is the best-fitting linear function for this binned data, by least-squares minimisation. We find that using all points individually gives nearly identical results, except when there are clear outliers, in which case our method results in visually better fits.

This procedure is repeated for all mass bins (i.e. not only the bins at $10^{13.5}$ and $10^{14.5}\mathrm{M_\odot}$ of Figures~\ref{fig:second_property_137} and \ref{fig:second_property_147}), resulting in mass-dependent values for the normalisation $\alpha$ and slope $\beta$ of the linear fit to $\delta \log_{10} L_{\rm X}(\delta \mathcal{Y})$ (Eq.~\ref{eq:fit_second_property_log}) for each property. As discussed in Section~\ref{methods:partial_correlation}, the normalisation and slope are fit with a third-order polynomial as a function of mass for each second property (Eq.~\ref{eq:fit_second_property_log_mass}). 

All relations between $\delta \log_{10} L_{\rm X}$ and a second property $\delta \mathcal{Y}$ show a large amount of scatter. At a fixed value of the second property, individual haloes can deviate from the best fit by up to $0.6 ~ \mathrm{dex}$, with a typical scatter of $0.2 ~ \mathrm{dex}$.

Tables~\ref{tab:best_alphas} and \ref{tab:best_betas} give the best fitting parameter values for the eight properties, out of Table~\ref{tab:all_properties}, that contribute more than $0.01~\mathrm{dex}$ correction to the scatter. Fig.~\ref{fig:betas_mass_dependence} shows the mass dependence of the Pearson-rank correlation coefficient of the six properties shown in Figures~\ref{fig:second_property_137} and \ref{fig:second_property_147}. Notably, for many quantities there is a peak, showing the mass scales where they correlate most strongly with the X-ray signal. We remark that the amplitude of the correlation does not translate directly into a reduction on the scatter in the X-ray luminosity. For that, the fitted slopes described above are multiplied with the deviation from a scaling relation. For quantities with larger scatter around their scaling relation, the same slope will result in a larger average correction.

\begin{table}
\caption{Best-fit parameters for the 8 most important secondary cluster properties -- out of all the properties listed in Table~\ref{tab:all_properties} -- that can be used to reduce the scatter of the $L_{\rm X}$ -- $M_{\rm 500c}$ relation according to Eqs.~\ref{eq:fit_second_property_log} and \ref{eq:correction_second_property_log}. Here we provide the values for the $c_1$, ..., $c_4$ parameters for $\alpha$ of Eq.~\ref{eq:fit_second_property_log_mass}, for FLAMINGO at $z=0$.}
\centering
\begin{tabular}{lrrrr}\hline
$\mathcal{Y}$ & $c_1|_\alpha$ & $c_2|_\alpha$ & $c_3|_\alpha$ & $c_4|_\alpha$ \\ \hline
$f_{\rm gas}$ & 20.81 & -4.60 & 0.34 & -0.01 \\ 
Y & -30.78 & 6.34 & -0.43 & 0.01 \\ 
$E_{\rm kin} / E_{\rm therm}$ & -11.64 & 2.43 & -0.17 & 0.00 \\ 
$T_{\rm gas}$ & -27.82 & 5.87 & -0.41 & 0.01 \\ 
$t_*$ & -18.51 & 3.83 & -0.26 & 0.01 \\ 
$M_{\rm MMBH}$ & -34.11 & 7.07 & -0.49 & 0.01 \\ 
$f_{\rm sat}$ & -22.10 & 4.69 & -0.33 & 0.01 \\ 
$c_{\rm DM}$ & -48.31 & 10.30 & -0.73 & 0.02 \\ 
\hline
\end{tabular}\label{tab:best_alphas}
\end{table} 

\begin{table}
\caption{Same as Table~\ref{tab:best_alphas} but for the $c_1$ through $c_4$ parameters for $\beta$ of Eq.~\ref{eq:fit_second_property_log_mass}.}
\centering
\begin{tabular}{lrrrr}\hline
$\mathcal{Y}$ & $c_1|_\beta$ & $c_2|_\beta$ & $c_3|_\beta$ & $c_4|_\beta$ \\ \hline
$f_{\rm gas}$ & 3348.90 & -727.49 & 52.58 & -1.26 \\ 
Y & -1222.22 & 250.07 & -17.00 & 0.38 \\ 
$E_{\rm kin} / E_{\rm therm}$ & 877.43 & -184.12 & 12.85 & -0.30 \\ 
$T_{\rm gas}$ & 7837.13 & -1631.94 & 112.99 & -2.60 \\ 
$t_*$ & -5275.32 & 1077.48 & -73.20 & 1.65 \\ 
$M_{\rm MMBH}$ & -1460.87 & 307.95 & -21.61 & 0.50 \\ 
$f_{\rm sat}$ & 283.69 & -58.35 & 3.99 & -0.09 \\ 
$c_{\rm DM}$ & -355.54 & 74.08 & -5.15 & 0.12 \\ 
\hline
\end{tabular}\label{tab:best_betas}
\end{table} 

\begin{figure*}
    \centering
    \includegraphics[width=0.45\linewidth]{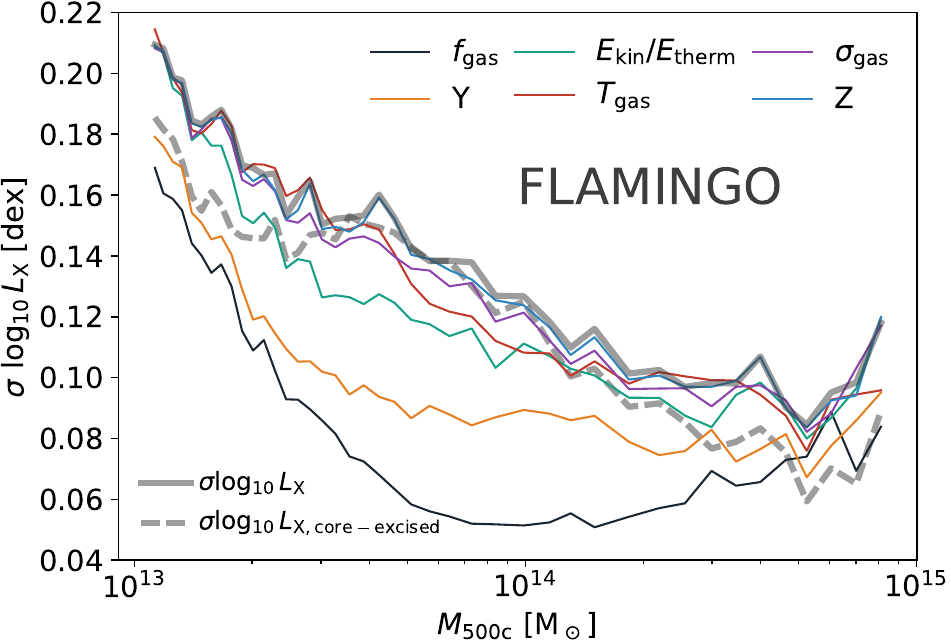}
    \includegraphics[width=0.45\linewidth]{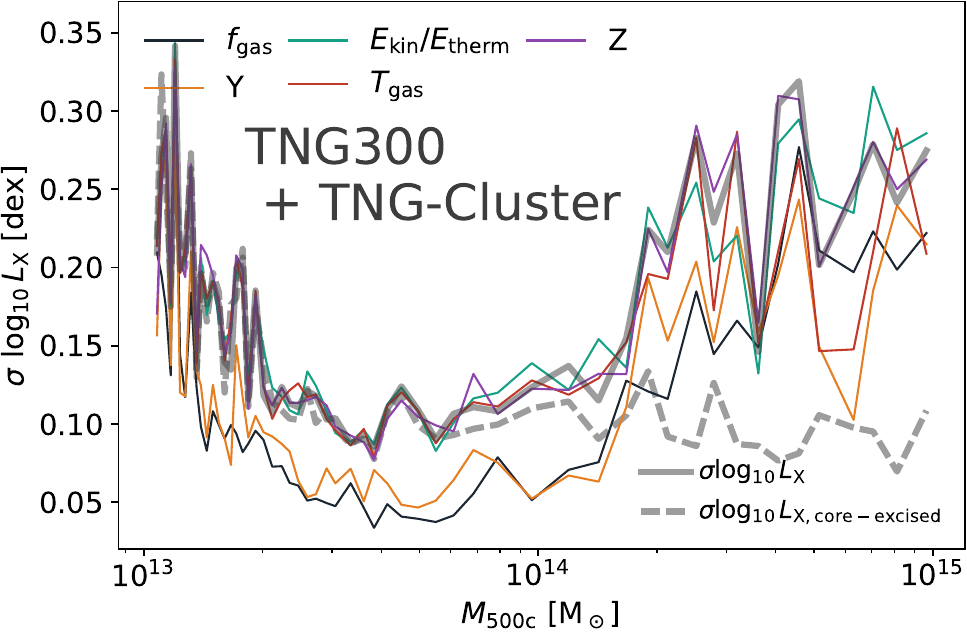}
    \includegraphics[width=0.45\linewidth]{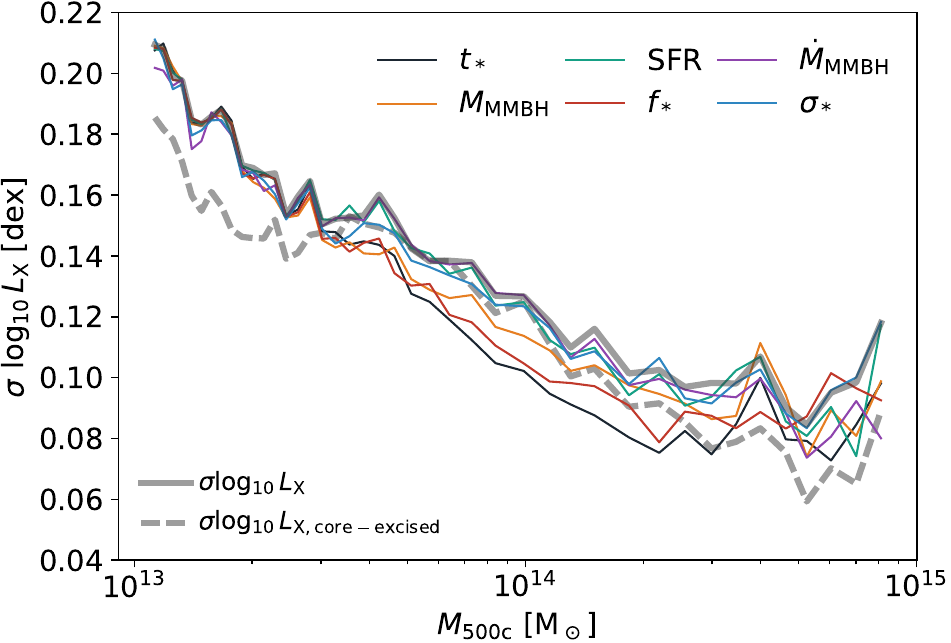}
    \includegraphics[width=0.45\linewidth]{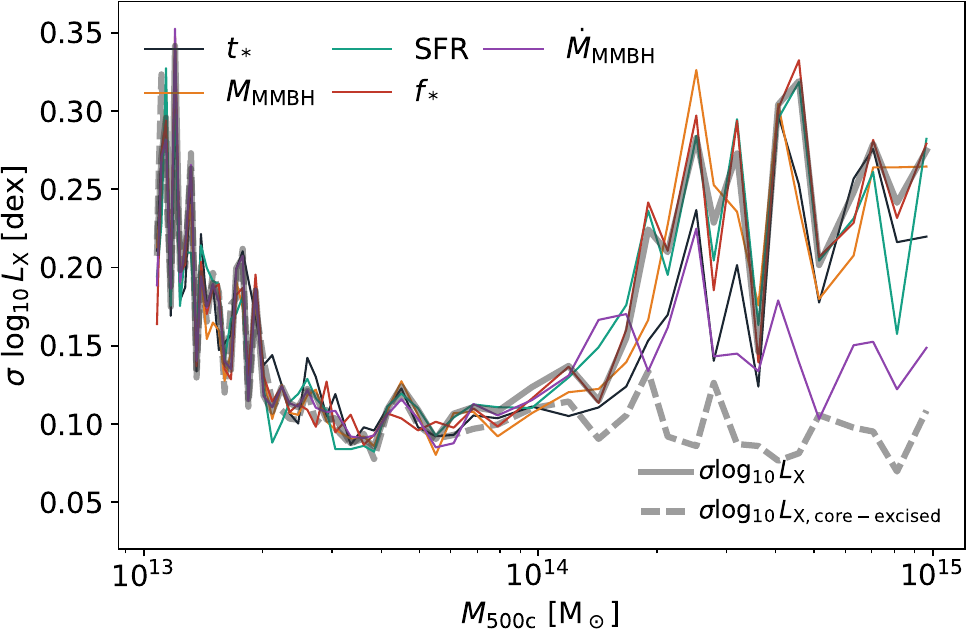}
    \includegraphics[width=0.45\linewidth]{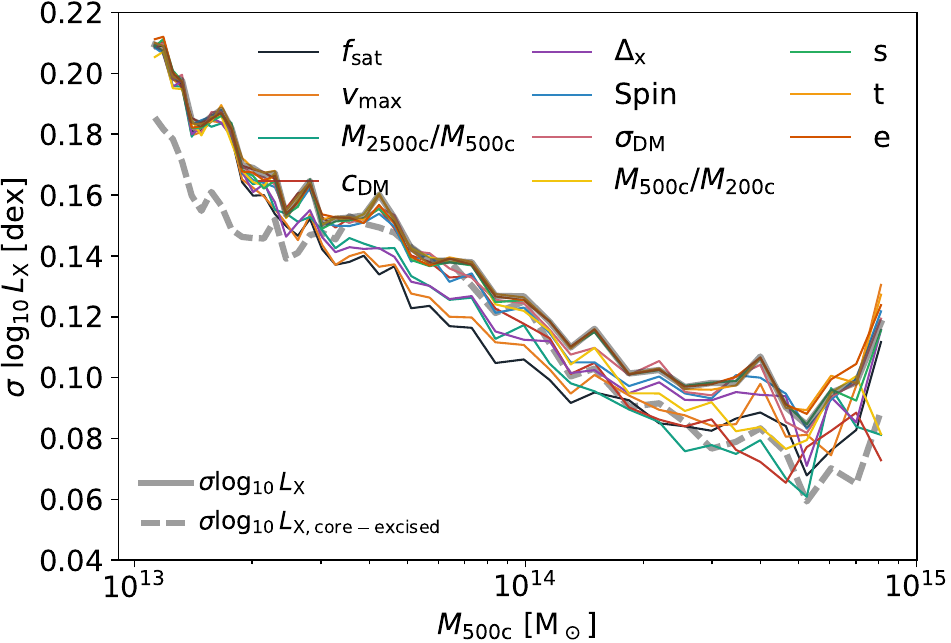}
     \includegraphics[width=0.45\linewidth]{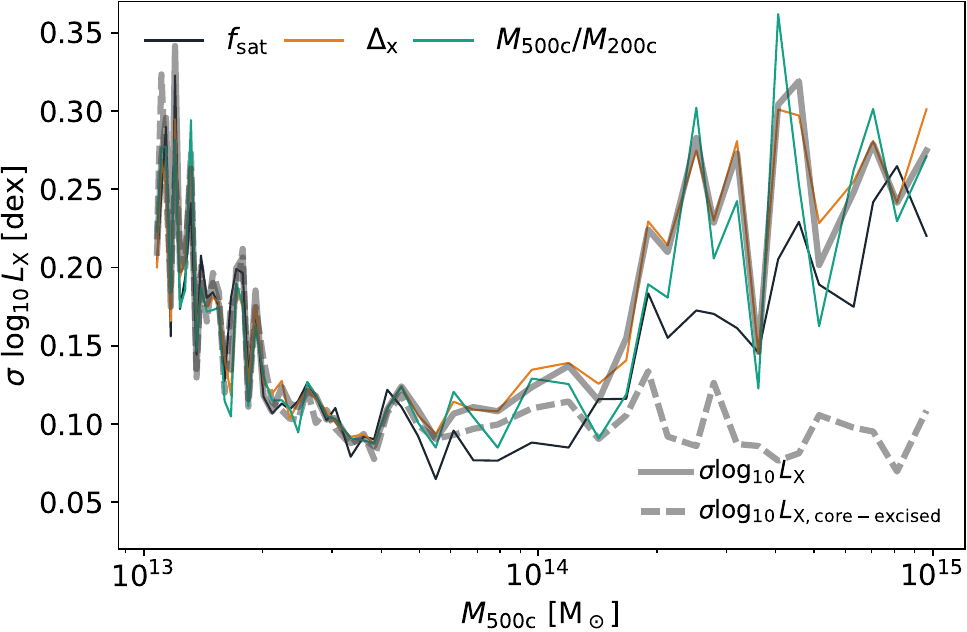}
    
    \caption{Reduction of the scatter in the $L_{\rm X}$ -- $M_{\rm 500c}$ scaling relation for the FLAMINGO L1m9 (left column) and TNG300+TNG-Cluster (right column) simulations at $z=0$, after accounting for the mass-dependent correlation between $L_{\rm X}$ and a second cluster property following Eq.~\ref{eq:correction_second_property_log}. The thick gray solid curve shows the intrinsic scatter without accounting for any other properties, the thick gray dashed curve shows the same but for core-excised X-ray luminosity. Thin solid coloured curves show the mass-dependent reduced scatter that can be obtained by knowing the value of another cluster property. The top panels show the scatter in $L_{\rm X}$ after accounting for correlations with ICM properties, the middle panel for galaxy properties, and the bottom panel for dark matter properties. Especially for $10^{13.5}-10^{14.5}~\mathrm{M_\odot}$ gas and baryonic properties yield larger reductions in  scatter than core-excision.}
    \label{fig:scatter_reduction_6panel}
\end{figure*}

\subsection{Reducing the scatter using a second property} \label{sec:reduce_scatter}
We now use the mass-dependent fits between the deviation from the X-ray luminosity -- halo mass relation and the properties in Table~\ref{tab:all_properties} to interpret the scatter. Using Equation~\ref{eq:correction_second_property_log}, we account for the scatter using our fits from Section~\ref{sec:correlate_with_second}.

Fig.~\ref{fig:scatter_reduction_6panel} compares the original scatter with the reduced scatter of the $L_{\rm X}$ -- $M_{\rm 500c}$ relation after correcting for the mass-dependent correlation between $\delta \log_{10} L_{\rm X}$ and a second property $\delta \mathcal{Y}$. The figure is split into three rows, which, following Table~\ref{tab:all_properties}, each show results from one type of second property. We report results for FLAMINGO in the left panels and for TNG300+TNG-Cluster on the right, but we discuss the latter in more detail later on. We focus on $z=0$.

The top row focuses on ICM properties. Here, we expect strong correlations, as the amount of gas directly affects the X-ray luminosity. Furthermore, bremsstrahlung dominates the X-ray luminosity for the massive haloes, and because it scales with the square root of the gas temperature, we also expect a correlation with $\delta T_{\rm gas}$. At low masses metal lines contribute to the X-ray luminosity and the expectation is not clear cut. The top row indeed confirms the expectation of strong correlations with ICM properties. Accounting for the gas fraction results in a strong reduction of the scatter, in both simulation models. For FLAMINGO (left panels), at $M_{\rm 500c} = 10^{14}~\mathrm{M_\odot}$, the scatter reduces from $0.15~\mathrm{dex}$ to $0.05~\mathrm{dex}$ (63\% reduction). For both simulation suites, we see a similarly strong reduction for Compton-Y, which itself correlates directly with gas fraction. In FLAMINGO, the ratio between cluster kinetic and thermal energies ($\delta [E_{\rm kin} / E_{\rm therm}]$) is effective at lower masses $10^{13.1} - 10^{14.0}~\mathrm{M_\odot}$, resulting in an $0.03 ~ \mathrm{dex}$ reduction in scatter. Gas temperature only results in a minor reduction of at most $0.02~\mathrm{dex}$ between $10^{13.5} - 10^{14.3}~\mathrm{M_\odot}$.

The middle row shows the scatter reduction when accounting for galaxy stellar and SMBH properties. Since none directly contribute to the X-ray luminosity, we expect smaller reductions in scatter. Nonetheless, in FLAMINGO, accounting for the most massive SMBH mass ($\delta M_{\rm MMBH}$) results in a similar reduction as the gas temperature. The mass-weighted mean stellar age ($\delta t_\star$) of the BCG gives a larger reduction but mostly towards larger masses. Between $10^{13.5} - 10^{14.7}~\mathrm{M_\odot}$ it reduces the scatter by $0.03 ~ \mathrm{dex}$. In TNG300+TNG-Cluster, the most massive SMBH accretion rate ($\delta \dot{M_{\rm MMBH}}$) results in a large reduction in scatter at the high-mass end, almost comparable to the core-excised result. None of the other properties result in a reduction of scatter of $L_{\rm X}$.

Finally, the bottom row uses correlations with dark matter halo properties to reduce the scatter. The deviation from the satellite fraction scaling relation ($\delta f_{\rm sat}$), and maximum circular velocity scaling relation ($\delta v_{\rm max}$) in FLAMINGO both offer an $0.02 ~ \mathrm{dex}$ reduction in scatter between $10^{13.5} - 10^{14.5}~\mathrm{M_\odot}$. We see a similar trend with satellite fraction in TNG300+TNG+Cluster. The Centre-of-mass -- Centre-of-potential offset deviation ($\delta \Delta_{\rm x}$) and $\delta [M_{\rm 2500c} / M_{\rm 500c}]$ offer a weaker reduction of $0.01 ~ \mathrm{dex}$ up to $10^{14.25}~\mathrm{M_\odot}$ in FLAMINGO, but has no effect in TNG300+TNG+Cluster. Above $10^{14.5}~\mathrm{M_\odot}$, the mass ratio $\delta [M_{\rm 2500c} / M_{\rm 500c}]$ and the concentration ($\delta c_{\rm DM}$) give reductions of up to $0.04 ~ \mathrm{dex}$ in FLAMINGO.
We find that knowledge of the halo shape, dark matter velocity dispersion, and spin parameter barely reduce the scatter in the X-ray luminosity.

We highlight here that for a large part of the mass range, the reduction in scatter we obtain is significantly larger than what can be achieved through core-excision. For FLAMINGO using gas properties results in a larger reduction for all masses $<5\times10^{14}~\mathrm{M_{\odot}}$, whereas for TNG300+TNG-Cluster the high-mass end shows enhanced scatter in the core region and core-excision is more effective for masses $>2\times 10^{14}~\mathrm{M_\odot}$. Especially in the regime of massive group or low-mass clusters, $\sim 10^{14}~\mathrm{M_\odot}$, using a second property results in a large reduction in the scatter, whereas core-excision has no impact on the scatter.

\begin{table}
    \centering
    \caption{Ratio of $\sigma \log_{10} L_{\rm X}$ after applying the correction from secondary properties with the intrinsic scatter, at 5 different halo masses (in $\mathrm{M_\odot}$) and  for the 8 most important secondary cluster properties, according to FLAMINGO at $z=0$. The ratio is the fraction of the scatter that remains unexplained after accounting for each property, separately. The mass where each property yields the largest reduction is highlighted in bold.}

    \begin{tabular}{l|l|l|l|l|l} 
        \hline
        $\mathcal{Y}$ & $10^{13.0}$ & $10^{13.5}$ & $10^{14.0}$ & $10^{14.5}$ & $10^{15.0}$ \\ \hline
        $f_{\rm gas}$ & 0.76 & 0.49 & \textbf{0.37} & 0.68 & 0.83  \\
        Y & 0.83 & \textbf{0.62} & 0.67 & 0.83 & 0.74 \\
        $E_{\rm kin} / E_{\rm therm}$ & 0.98 & \textbf{0.81} & 0.86 & 0.83 & 1.01 \\ 
        $T_{\rm gas}$ & 1.00 & 0.97 & \textbf{0.84} & 1.01 & 0.96 \\ 
        $t_*$ & 1.01 & 0.88 & \textbf{0.81} & 0.83 & 0.90 \\ 
        $M_{\rm MMBH}$ & 1.02 & 0.93 & 0.88 & \textbf{0.87} & 0.98 \\ 
        $f_{\rm sat}$ & 1.00 & 0.88 & \textbf{0.81} & 0.83 & 0.90 \\ 
        $c_{\rm DM}$ & 1.01 & 0.98 & 0.92 & 0.87 & \textbf{0.74} \\  \hline
    \end{tabular}
    \label{tab:strongest_scatter_reductions_mass}
\end{table}

\begin{table}
    \centering
    \caption{As Table~\ref{tab:strongest_scatter_reductions_mass} but for the 7 most important secondary properties according to TNG300+TNG-Cluster.}

    \begin{tabular}{l|l|l|l|l|l} 
        \hline
        $\mathcal{Y}$ & $10^{13.0}$ & $10^{13.5}$ & $10^{14.0}$ & $10^{14.5}$ & $10^{15.0}$ \\ \hline
        $f_{\rm gas}$ & 1.00 & 0.43 & \textbf{0.38} & 0.53 & 0.76  \\
        Y & 0.70 & 0.57 & \textbf{0.39} & 0.78 & 0.73 \\
        $E_{\rm kin} / E_{\rm therm}$ & 1.04 & 0.89 & 1.15 & \textbf{0.75} & 1.05 \\ 
        $T_{\rm gas}$ & 1.03 & 0.92 & 1.04 & 1.07 & \textbf{0.70} \\ 
        $t_*$ & 1.00 & 1.16 & 0.89 & \textbf{0.67} & 0.75 \\ 
        $M_{\rm MMBH}$ & 1.02 & 0.92 & 0.85 & \textbf{0.82} & 0.95 \\ 
        $f_{\rm sat}$ & 0.96 & 1.07 & 0.68 & \textbf{0.51} & 0.75 \\ \hline
    \end{tabular}
    \label{tab:strongest_scatter_reductions_mass_tng}
\end{table}

We also explore the maximum reduction in the scatter in $L_{\rm X}$ that can be achieved by combining several secondary properties. To this end, per category (ICM, galaxy, dark matter), we first apply the second property resulting in the largest reduction in scatter from Fig.~\ref{fig:scatter_reduction_6panel}. After that, we fit the residual of the scatter using the second most strongly correlated property, proceeding this way for all properties in each category. We find that after applying the first, strongest reduction, no other properties contribute a further individual or combined reduction in the scatter of more than  $0.01~\mathrm{dex}$. This leads us to conclude that the strongest reduction seen in each category ($f_{\rm gas}$, $t_\star$, and $f_{\rm sat}$) correlates with the other properties in the same category that show reductions in Fig.~\ref{fig:scatter_reduction_6panel}, since none of them contribute a further, independent reduction.

A summary of the strongest reductions in scatter is in Table~\ref{tab:strongest_scatter_reductions_mass} (FLAMINGO) and Table~\ref{tab:strongest_scatter_reductions_mass_tng} (TNG300+TNG-Cluster). Here, again, it is clear that the ICM properties ($f_{\rm gas}$ and Compton-$Y$) yield the strongest reduction in scatter. It also becomes clear that those properties are most important at the boundary between group and cluster mass scales. Other properties in the table show a smaller impact on the scatter, and are effective at higher masses. The most extreme case is the halo concentration, $c_{\rm DM}$, which yields the largest scatter reduction for the most massive clusters ($M_{\rm 500c} = 10^{15.0}~\mathrm{M_\odot}$). We note that in TNG300+TNG-Cluster the maximum reduction of the scatter typically happens at higher mass.

In certain cases correlations seemingly exists when judged by eye, but the scatter of these correlations is too large for them to serve as a useful correction. This is the case, for example, for the concentration at high halo mass. The bottom right panel of Fig.~\ref{fig:second_property_147} shows that even for haloes with $-0.5~\mathrm{dex}$ offset from the median concentration, there are still objects with a positive offset from the X-ray luminosity scaling relation. The correction, determined by the mean in the bin, will in such cases increase the offset from the $L_{\rm X}$ scaling relation, simply because the distribution of $\delta c_{\rm DM}$ - $\delta L_{\rm X}$ is not Gaussian. This is also seen in other properties explored in this work.

More properties can be thought of which could correlate with the scatter in the X-ray luminosity at fixed halo mass. One of those that has recently garnered interest is the formation time of haloes. We have checked for correlation with $z_{1/2}$, but do not find any in FLAMINGO. This is most likely due to the significant scatter in the $\delta L_{\rm X}$ - $\delta z_{1/2}$ relation, such that, as noted above, the existence of a correlation can still result in a negligible correction.

\begin{figure}
    \centering
    \includegraphics[width=\linewidth]{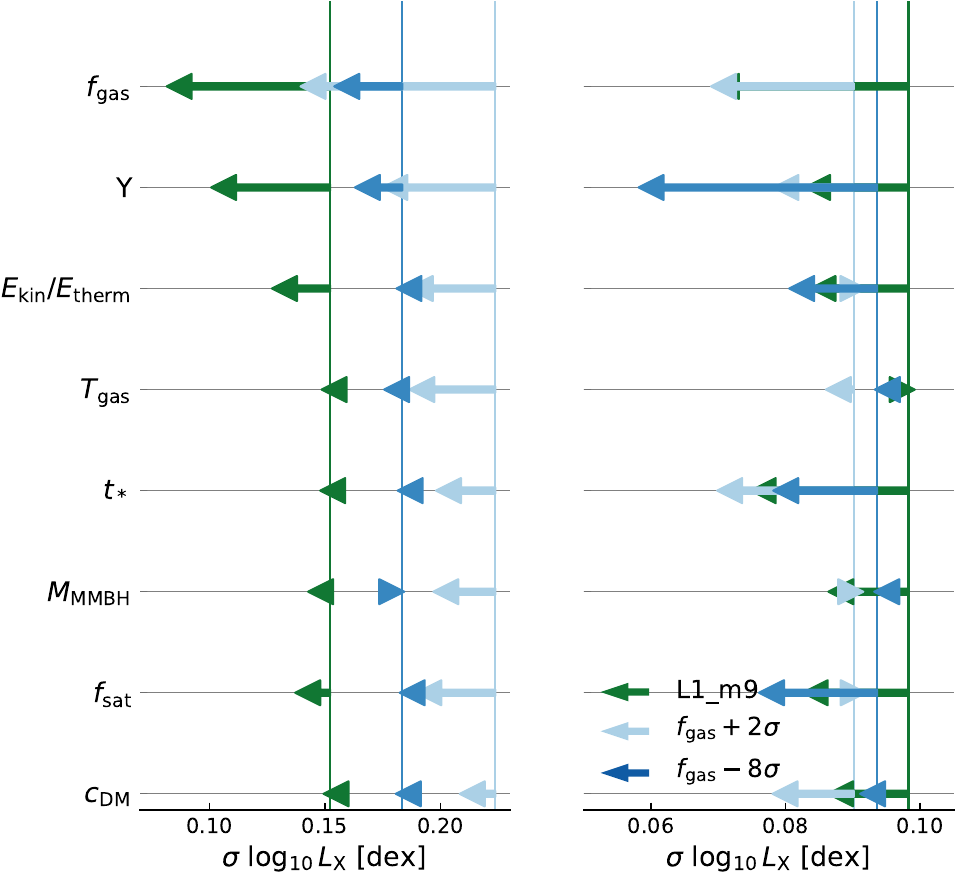}
    \caption{The scatter in the $L_{\rm X}$ -- $M_{\rm 500c}$ relation at $z=0$ for three different FLAMINGO galaxy formation-model variations, L1\_m9 (fiducial, green), $f_{\rm gas}+2\sigma$ (light blue), $f_{\rm gas}-8\sigma$ (dark blue). The left panel shows results for $M_{\rm 500c} = 10^{13.5}~\mathrm{M_\odot}$, the right panel for $M_{\rm 500c} = 10^{14.5}~\mathrm{M_\odot}$. The horizontal axis ranges differ between the two panels to highlight the relevant scatter values for both masses. Solid coloured vertical lines indicate the level of scatter for the different simulations, and arrow heads point towards the scatter after applying a correction for the indicated quantity. At $M_{\rm 500c} = 10^{13.5}~\mathrm{M_\odot}$, $f_{\rm gas}+2\sigma$ has the most scatter, with a larger number of secondary properties resulting in reduced scatter. At $M_{\rm 500c} = 10^{14.5}~\mathrm{M_\odot}$ all model variations have similar scatter. It is clear from this figure that $f_{\rm gas}$ and Y are always the most important properties.}
    \label{fig:gas_fraction_scatter_comparison}
\end{figure}

\subsection{Comparison between FLAMINGO feedback models} \label{sec:gas_fraction}
The feedback variations within the FLAMINGO suite offer a unique opportunity to statistically assess the galaxy group and cluster population. The highest cluster gas fraction model variation ($f_{\rm gas} + 2\sigma$) has significantly weaker AGN feedback. 
In this weaker feedback model, the relative importance of processes besides AGN feedback in the halo energy balance could potentially change. Conversely, in the stronger feedback model, the AGN has deposited significantly more energy in the halo, leading not only to lower gas fractions, but also higher temperatures. Hence, the scatter in the X-ray luminosity and the correlation of different halo properties with this scatter could change between the feedback model variations.

We see this in Figure~ ~\ref{fig:gas_fraction_scatter_comparison}, based on $z=0$ results, where $f_{\rm gas} + 2\sigma$ has $0.08~\mathrm{dex}$ more scatter around $M_{\rm 500c} = 10^{13.5}~\mathrm{M_\odot}$ than L1\_m9.
The $f_{\rm gas} - 8\sigma$ model falls in between the fiducial and high gas fraction model, showing that the scatter is not monotonic with gas fraction.
The increased scatter for $f_{\rm gas} + 2\sigma$ could be explained because the AGN feedback, in this model, is not powerful enough to dominate the halo and other secondary properties become important. This is illustrated by the stellar age, most massive SMBH mass, satellite mass fraction, and maximum circular velocity all leading to larger reductions of the scatter in the $L_{\rm X} - M_{\rm 500c}$ relation for $f_{\rm gas} + 2\sigma$, compared to the lower gas fraction models.
Also notable, at this mass the lowest gas fraction model allows very small corrections to the scatter. The deviation in gas fraction and Compton-y signal become significantly less informative than for the fiducial model.

At higher masses ($M_{\rm 500c} = 10^{14.5}~\mathrm{M_\odot}$), the scatter for all model variations is equal within $0.01~\mathrm{dex}$. There is, however, a large spread in which second properties result in a reduction in the scatter. For all three variations, accounting for the deviation in gas fraction results in a large reduction. For the lowest gas fraction model ($f_{\rm gas} - 8\sigma$), accounting for the deviation in Compton-Y signal actually results in a larger reduction than the deviation from the gas fraction scaling relation.
The mass-weighted mean stellar age ($t_\star$), contributes strongly for all three models, and gives a $0.02-0.03~\mathrm{dex}$ reduction in the scatter. The satellite mass fraction contributes slightly for L1\_m9 and $f_{\rm gas}-8\sigma$, but not for $f_{\rm gas}+2\sigma$. 

\begin{figure}
    \centering
    \includegraphics[width=\linewidth]{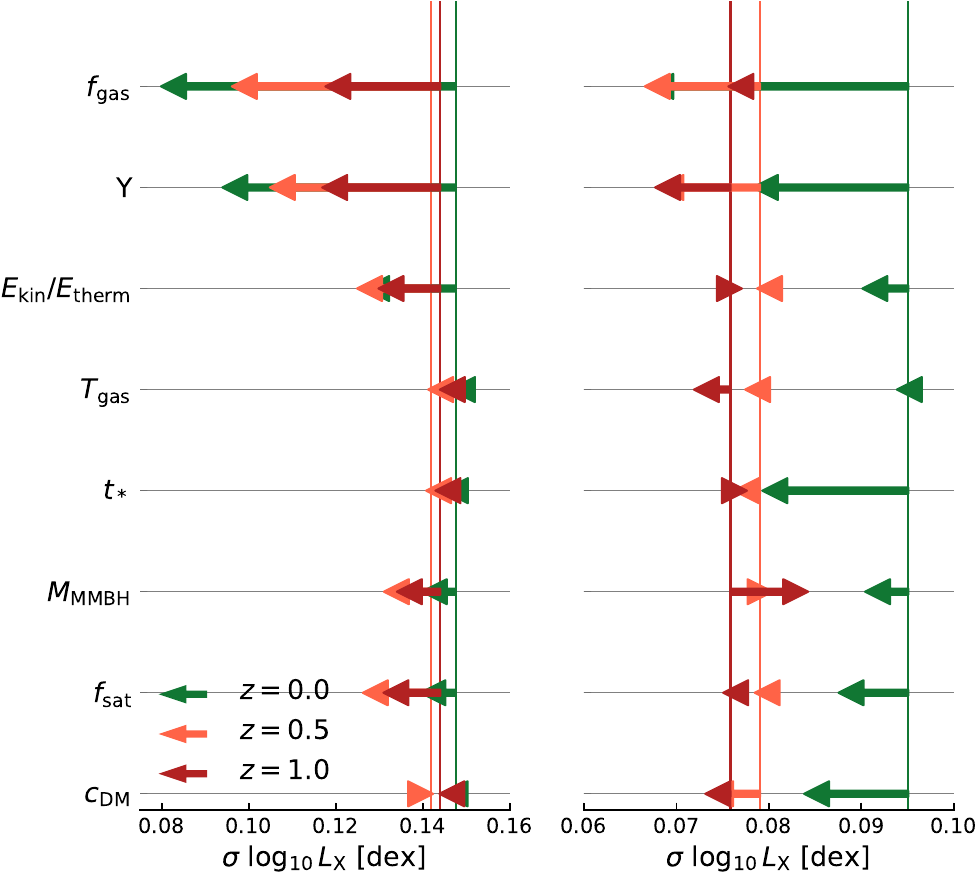}
    \caption{As Fig.~\ref{fig:gas_fraction_scatter_comparison}, but for the fiducial FLAMINGO L2p8 simulation only and for three different redshifts: $z = 0$ (green), $z = 0.5$ (orange), $z = 1$ (red). $z = 0$ has a higher scatter at $M_{\rm 500c} = 10^{14.5}~\mathrm{M_\odot}$ than the other redshifts. At $z=0$ accounting for the deviation in $t_\star$ and $c_{\rm DM}$ results in a strong reduction in the scatter, unlike at higher redshifts. At $M_{\rm 500c} = 10^{13.5}~\mathrm{M_\odot}$ all redshifts have similar scatter, but the maximum reductions to the scatter decrease with increasing redshift.}
    \label{fig:redshift_scatter_comparison}
\end{figure}

\subsection{Redshift evolution of the scatter} \label{sec:redshift}
What about the redshift evolution of the scatter and whether contributors to the scatter reduction change with cosmic time? We make use of the $(2.8~\mathrm{Gpc})^3$ FLAMINGO volume at $z = 0, 0.5$ and $1$, selecting all haloes with $M_{\mathrm{500c}} > 10^{13}~\mathrm{M_\odot}$. For all redshifts there are $> 10^4$ haloes with mass $10^{13.45} - 10^{13.55}~\mathrm{M_\odot}$; between $10^{14.45}-10^{14.55}~\mathrm{M_\odot}$ there are 6981 objects at $z=0$, and 372 objects at $z=1$.

As shown in Fig.~\ref{fig:scatter_redshift_FLAMINGO_TNG}, for the group scale ($M_{\rm 500c} = 10^{13.5}~\mathrm{M_\odot}$) the scatter does not change with redshift (only the slope of the scatter), but for the cluster scale ($M_{\rm 500c} = 10^{14.5}~\mathrm{M_\odot}$), there is a pronounced decrease of $0.02~\mathrm{dex}$ from $z=0$ to $z = 0.5$, though no further decrease to $z=1$. 

Fig.~\ref{fig:redshift_scatter_comparison} shows that the most important properties that contribute to a reduction in the scatter are the same for all redshifts. The deviations from the gas fraction and Compton-Y scaling relations always dominate. However, they become increasingly less informative (i.e. giving smaller reductions) at higher redshift. For galaxy groups (left) the reduction after accounting for the gas fraction deviation decreases from $0.06~\mathrm{dex}$ at $z=0$, to $0.02~\mathrm{dex}$ at $z = 1$. Similarly, the effect of accounting for deviations from the Compton-Y scaling relation decreases from $0.04~\mathrm{dex}$ to $0.02~\mathrm{dex}$. For clusters (right) the same effect is seen, the impact of both deviations in gas fraction and Compton-Y signal decreases from $0.03~\mathrm{dex}$ to $0.01~\mathrm{dex}$ between $z=0$ and $z=1$. Evidently, at high redshift, more of the scatter is not explained by any of the properties studied here.

For galaxy clusters at $z=0$, other properties such as the deviation in the BCG mean stellar age ($\delta t_\star$) and the halo concentration ($\delta c_{\rm DM}$) contribute $0.01-0.02 ~\mathrm{dex}$, but at $z=0.5$ and $z=1$ only the deviation in gas fraction and Compton-Y signal remain as non-negligible contributors. This hints at an alignment between the cluster core and global X-ray luminosity at $z=0$, which is not yet present at higher redshift.

In short, the main contributors to the scatter do not change with redshift, though the magnitude of the contributions of the properties does change.

\subsection{Comparison with the TNG300+TNG-Cluster simulations} \label{sec:tng_comparison}
We finally discuss the comparison between FLAMINGO and the TNG300+TNG-Cluster simulations. Because there is no way to remove gas cells recently heated by AGN feedback in TNG300+TNG-Cluster analogous to the procedure we apply in FLAMINGO, we choose to use core-excised X-ray luminosities for both simulations when comparing. We excise the region $r < 0.15~r_{\rm 500c}$, which, apart from being a common observational choice, safely excludes all recently heated particles in FLAMINGO. This exclusion is motivated by the sharp upturn in the scatter observed in TNG300+TNG-Cluster above $M_{\rm 500c}=10^{14}~\mathrm{M_\odot}$ when the core is not removed (see Fig. \ref{fig:scatter_redshift_FLAMINGO_TNG}).

\begin{figure}
    \centering
    \includegraphics[width=\linewidth]{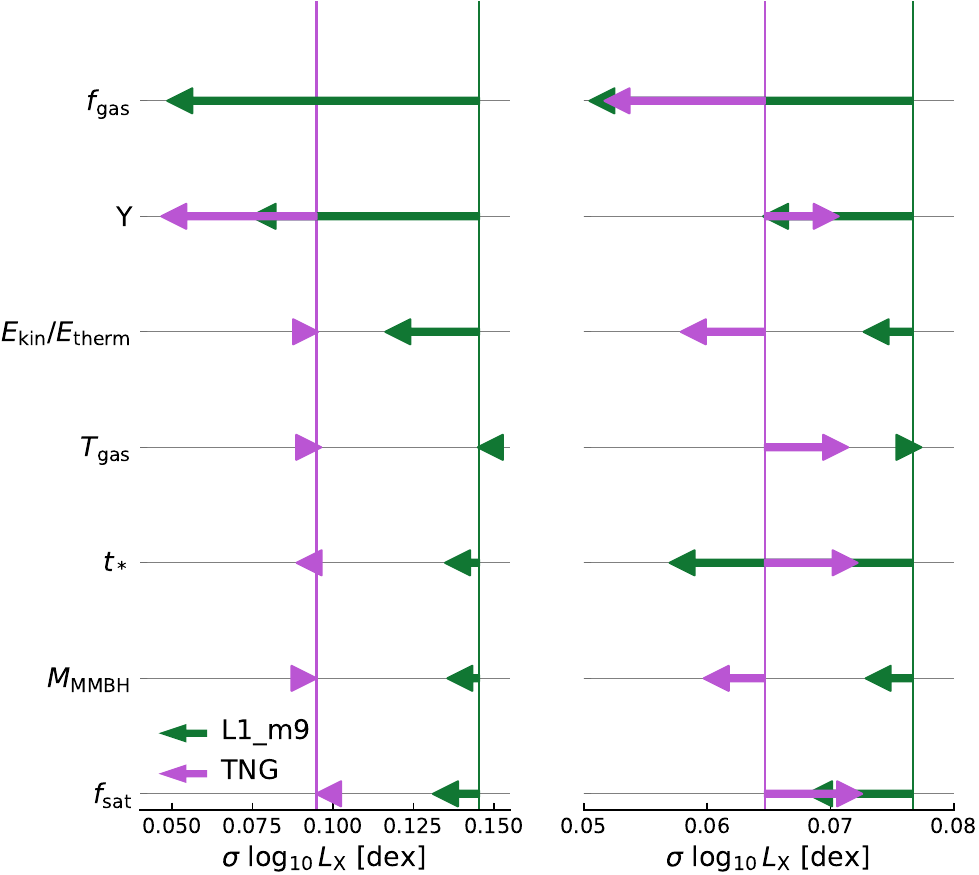}
    \caption{As Fig.~\ref{fig:gas_fraction_scatter_comparison}, but now for core-excised X-ray luminosity, comparing the strongest contributors to scatter reduction for L1\_m9 (green) and TNG300+TNG-Cluster (violet). Groups and clusters from TNG300+TNG-Cluster display less scatter at both masses than FLAMINGO. At $M_{\rm 500c} = 10^{13.5}~\mathrm{M_\odot}$ both simulations show strong reductions in the scatter when accounting for $f_{\rm gas}$ and Compton-$Y$. For $M_{\rm 500c} = 10^{14.5}~\mathrm{M_\odot}$, only the gas fraction shows a coherent signal across both simulations.}
    \label{fig:flamingo_tng_comparison}
\end{figure}

Fig.~\ref{fig:flamingo_tng_comparison} shows that, despite the completely independent feedback choices and subgrid methods, both simulations yield similar scatter in the X-ray luminosity scaling relation, especially at the cluster scale. At $M_{\rm 500c} = 10^{13.5}~\mathrm{M_\odot}$, where FLAMINGO has a scatter of around $0.15~\mathrm{dex}$, TNG300+TNG-Cluster has $0.05~\mathrm{dex}$ less scatter. As a result, the TNG300+TNG-Cluster scatter is much less mass dependent. Both simulations show that accounting for the deviation from the gas fraction and Compton Y scaling relations result in a strong reduction in the scatter at the lower mass. FLAMINGO also shows a dependence on the dynamical state through $\delta [E_{\rm kin}/E_{\rm therm}]$, which is not seen for TNG300+TNG+Cluster. At the higher mass, FLAMINGO shows a dependence of the scatter on the deviation in stellar age $\delta t_\star$, which is absent in TNG300+TNG+Cluster. Conversely, TNG300+TNG-Cluster shows a weak dependence on the star formation rate and stellar mass fraction, unlike FLAMINGO. Overall, the levels of scatter at the high-mass end are remarkably similar.

\section{Conclusions} \label{sec:conclusions}
We study the mass dependence and redshift evolution of the scatter in the X-ray luminosity -- halo mass relation of galaxy groups and clusters ($M_{\rm 500c} > 10^{13} ~ \mathrm{M_\odot}$, $z<2$) from the FLAMINGO and TNG300+TNG-Cluster cosmological hydrodynamical simulations for galaxy cluster physics. Thanks to their large volumes compared to previous cluster simulations, particularly for FLAMINGO, we are equipped with abundant statistics also at the highest-end of the halo mass function. We can hence robustly quantify the scatter in the $L_{\rm X}$ -- $M_{\rm 500c}$ relation and provide analytic fitting functions for its redshift evolution. We also study how accounting for other halo properties can reduce the $L_{\rm X}$ -- $M_{\rm 500c}$ scatter. To this end, we fit linear relations between residuals of the X-ray luminosity -- halo mass relation and the residuals of the scaling relation with halo mass of a second property. 
Fitting functions and parameters are provided for all FLAMINGO variations and the combination of TNG300 and TNG-Cluster.

Our main findings are: 
\begin{itemize}
    \item The scatter about the median $L_{\rm X}$ -- $M_{\rm 500c}$ relation in FLAMINGO is 0.22 $\mathrm{dex}$ at $M_{\rm 500c} = 10^{13}~\mathrm{M_\odot}$, decreasing to 0.1 $\mathrm{dex}$ around $10^{14}~\mathrm{M_\odot}$, after which it remains constant with mass. For TNG300+TNG-Cluster, the scatter is larger at $10^{13}~\mathrm{M_\odot}$ at 0.27 $\mathrm{dex}$, sharply dropping to 0.1 $\mathrm{dex}$ around $10^{13.5}~\mathrm{M_\odot}$, after which it increases again to 0.25 $\mathrm{dex}$ around $10^{14.5}~\mathrm{M_\odot}$. This increase in scatter at high mass in TNG300+TNG-Cluster is driven by the core region, and core-excision results in a constant scatter at the highest masses that is similar in magnitude to FLAMINGO (Fig.~\ref{fig:scatter_redshift_FLAMINGO_TNG}).
    \item The mass-dependence of the scatter is well fit with a broken power law, and the redshift evolution of the power-law parameters with a quadratic function. This best fit recovers the FLAMINGO scatter within $0.01~\mathrm{dex}$ between $10^{13}$-$10^{15.5}~\mathrm{M_\odot}$ and $z=0-2$. Due to the much smaller volume, there is more bin-to-bin variation in TNG300+TNG-Cluster, but the scatter is still recovered within $0.05~\mathrm{dex}$ in the same mass and redshift range (Fig.~\ref{fig:scatter_redshift_FLAMINGO_TNG}).
    \item The deviation from the $L_{\rm X}$--$M_{\rm 500c}$ relation correlates strongly with the gas fraction at all masses, in both FLAMINGO and TNG300+TNG-Cluster. For the former, at the cluster-mass scale the scatter in $L_{\rm X}$--$M_{\rm 500c}$ also correlates with the stellar age of the BCG ($t_\star$) and the halo concentration ($c_{\rm DM}$) (Figures \ref{fig:second_property_137} and \ref {fig:second_property_147}).
    \item The scatter in $L_{\rm X}$ -- $M_{\rm 500c}$ can be strongly reduced using knowledge of the gas fraction, e.g. from $0.15~\mathrm{dex}$ to $0.05~\mathrm{dex}$ at $10^{14}~\mathrm{M_\odot}$ according to FLAMINGO. Accounting for the Compton Y signal yields a slightly smaller reduction, but with the same mass dependence. The measure of halo disturbance $E_{\rm kin} / E_{\rm therm}$ gives a small reduction of $0.02-0.03~\mathrm{dex}$ below $10^{14}~\mathrm{M_\odot}$ in FLAMINGO, but not in TNG300+TNG+Cluster. Other ICM quantities such as temperature and gas metallicity are not useful to minimize the scatter (Fig.~\ref{fig:scatter_reduction_6panel}).
    \item In FLAMINGO, accounting for the stellar age of the BCG can reduce the scatter in $L_{\rm X}$ -- $M_{\rm 500c}$ by $0.03~\mathrm{dex}$ between $10^{13.5}$ and $10^{14.5}~\mathrm{M_\odot}$, and accounting for the mass of the most massive SMBH results in a $0.01~\mathrm{dex}$ reduction around $10^{13.75}~\mathrm{M_\odot}$. In FLAMINGO no other stellar or SMBH properties correlate with the scatter, but in TNG300+TNG-Cluster the accretion rate of the most massive SMBH correlates strongly for $M_{\rm 500c} > 10^{14.5}~\mathrm{M_\odot}$ (Fig.~\ref{fig:scatter_reduction_6panel}).
    \item Multiple dark matter properties can explain $0.01-0.02~\mathrm{dex}$ of the scatter of $L_{\rm X}$ -- $M_{\rm 500c}$, with the strongest contributor on group scales being the satellite mass fraction ($f_{\rm sat}$), a measure of relaxedness, and at cluster mass scales the dark matter concentration ($c_{\rm DM}$) (Fig.~\ref{fig:scatter_reduction_6panel}).
    \item The FLAMINGO feedback variations all show similar correlations between the scatter in $L_{\rm X}$ -- $M_{\rm 500c}$ and secondary properties. The deviations in gas fraction and Compton Y signal always correlate strongest with the deviation from the X-ray luminosity scaling relation. For massive clusters the deviation in satellite mass fraction, stellar age of the BCG, and halo concentration have small contributions. For the weakest feedback model ($f_{\rm gas}+2\sigma$), at the group-mass scale, many other properties result in a small $0.01-0.02~\mathrm{dex}$ reduction in scatter, at the expense of the gas fraction being slightly less informative (Fig.~\ref{fig:gas_fraction_scatter_comparison}).
    \item The main contributors to the scatter remain the same with redshift (out to $z = 1$), though their contributions diminish with increasing redshift. The decrease in scatter when accounting for the gas fraction at $10^{13.5}~\mathrm{M_\odot}$ changes from $0.06~\mathrm{dex}$ at $z = 0$ to $0.02~\mathrm{dex}$ at $z = 1$ (Fig.~\ref{fig:redshift_scatter_comparison}).
    \item Though the absolute level of scatter is different between FLAMINGO and TNG300+TNG-Cluster, the scatter reduction due to various second properties is similar. The deviation in gas fraction dominates the reduction, followed by Compton Y. For clusters, FLAMINGO has a contribution to the scatter from the deviation in the BCG stellar age, which we do not see in TNG300+TNG-Cluster (Fig.~\ref{fig:flamingo_tng_comparison}).
\end{itemize}

Even when accounting for the properties that correlate most strongly with the X-ray luminosity, 40\% of the scatter remains unexplained at $M_{\rm 500c} = 10^{14}~\mathrm{M_\odot}$, which rises to 70\% at lower ($10^{13}~\mathrm{M_\odot}$) and higher ($10^{15}~\mathrm{M_\odot}$) masses. This could potentially be due to the stochastic nature of AGN feedback \citep{borrow_impact_2023}, but it remains unclear whether this can fully explain the scatter and the mass-dependence.

Between masses of $10^{13.5}$ and $10^{14.5}~\mathrm{M_\odot}$ the scatter is driven by processes outside the core ($0.15~R_{\rm 500c}$) region. In both FLAMINGO and TNG300+TNG-Cluster, core-excision does not result in a reduction of the $L_{\rm X}$ -- $M_{\rm 500c}$ scatter in this mass range. On the other hand, the methodology presented in this work provides a much more effective avenue to reduce the scatter in groups and clusters analyses, being most effective for massive groups and low-mass clusters, where the reduction can reach $60\%$.

This is the first systematic study into the origin of the scatter in the X-ray luminosity -- halo mass relation for groups and clusters. It has become feasible to do so thanks to the large sample size of massive haloes in the FLAMINGO simulation ($>2,000,000$ haloes above $10^{13}~\mathrm{M_\odot}$), with accurate calibration to halo gas fractions: this has made statistical methods that before could only be applied to galaxies accessible also for galaxy groups and clusters. By combining FLAMINGO with the outcome of the TNG300+TNG-Cluster simulations, we have shown that the scatter in the $L_{\rm X}$ -- $M_{\rm 500c}$ relation can be significantly reduced without having to reduce the sample size. 
The strongest independent measurement resulting in a reduction of the scatter in $L_{\rm X}$ -- $M_{\rm 500c}$ is that of the Compton Y signal, which can be measured from SZ surveys. We have shown that this is true across redshift, across variations of the same simulation model, and even across completely different simulation projects and hence galaxy-physics choices and implementations. Using such methods to reduce the scatter opens up the potential of tighter constraints on cosmological parameters from large X-ray surveys.

\section*{Acknowledgements}
We acknowledge support from research programme Athena 184.034.002 from the Dutch Research Council (NWO). JB and AP acknowledge funding from the
European Union (ERC, COSMIC-KEY, 101087822, PI: Pillepich). DN acknowledges funding from the Deutsche Forschungsgemeinschaft (DFG) through an Emmy Noether Research Group (grant number NE 2441/1-1).
This work used the DiRAC@Durham facility managed by the Institute for Computational Cosmology on behalf of the STFC DiRAC HPC Facility (www.dirac.ac.uk). The equipment was funded by BEIS capital funding via STFC capital grants ST/K00042X/1, ST/P002293/1, ST/R002371/1 and ST/S002502/1, Durham University and STFC operations grant ST/R000832/1. DiRAC is part of the National e-Infrastructure. This project has received funding from the European Research Council (ERC) under the European Union’s Horizon 2020 research and innovation programme (grant agreement No 769130).

\section*{Data Availability}
The data underlying the plots within this article are available on
reasonable request to the corresponding author. The FLAMINGO simulation data will eventually be made publicly available, though we note that the data volume (several petabytes) may prohibit us from simply placing the raw data on a server. In the meantime, people interested in using the simulations are encouraged to contact the corresponding author. 
All the IllustrisTNG simulations, including TNG300, and TNG-Cluster are publicly available and accessible at \url{www.tng-project.org/data}, as described in \cite{nelson_illustristng_2019}.



\bibliographystyle{mnras}
\bibliography{references} 




\appendix
\section{Cosmology dependence} \label{sec:cosmology}
\begin{figure*}
    \centering
    \includegraphics[width=\linewidth]{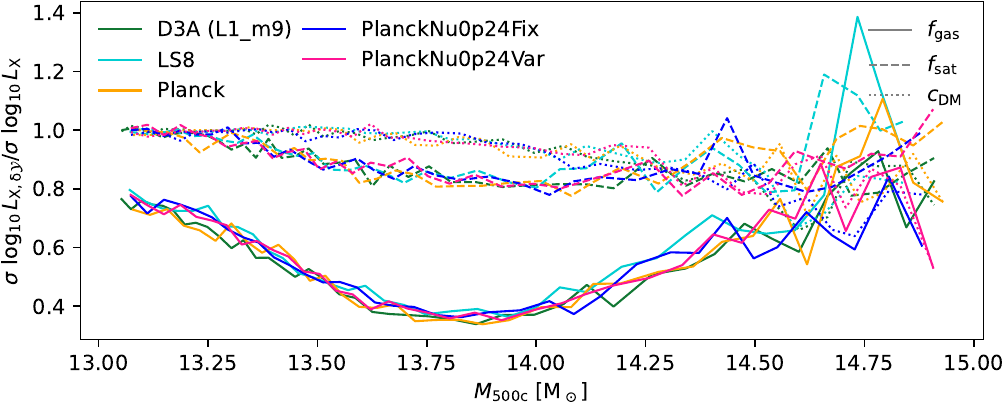}
    \caption{The fractional reduction in scatter in the X-ray luminosity -- halo mass relation when accounting for an additional physical quantity, in different background cosmologies. Different line styles indicate three additional quantities, colors show different background cosmologies. There is no difference in the reduction of scatter with a change in background cosmology.}
    \label{fig:scatter-reduction-cosmology}
\end{figure*}
We study the change in scatter about the median $L_{\rm X}-M_{\rm 500c}$ relation with differing cosmology, using the $(1~\mathrm{Gpc})^3$ cosmology variations within the FLAMINGO suite of simulations. For each simulation we carry out the analysis described in Section~\ref{methods:partial_correlation}. 
The default cosmology (DES year 3) is compared with a vanilla Planck cosmology, a lensing low $\sigma_8$ cosmology (LS8), and two cosmologies with a neutrino mass of $0.24~\mathrm{eV}$ (see Table 2 of \citet{schaye_flamingo_2023} for the parameters of each model).

Fig.~\ref{fig:scatter-reduction-cosmology} shows that the background cosmology has no impact on the contributions of different additional properties. We show the property yielding the strongest scatter reduction ($f_{\rm gas}$), and two properties which depend more on the history of the halo and hence might be cosmology dependent ($c_{\rm DM}$ and $f_{\rm sat}$). The reduction of the scatter due to all of these properties is independent of the background cosmology at all masses.

\section{Percentile or lognormal scatter} \label{sec:appendix_lognormal}
A common approach when measuring the scatter in scaling relations is to assume that the distribution of values at a fixed mass is lognormal. However, it has been shown that this is not necessarily true, especially at the group-mass scale \citep[$10^{13}~\mathrm{M_\odot}$;][]{kugel_flamingo_2024}. Here we quantify the mass-dependent difference in scatter between assuming a lognormal and measuring the interval between the 16$^{\rm th}$ and 84$^{\rm th}$ percentiles, where the latter is insensitive to the shape of the distribution.

\begin{figure}
    \centering
    \includegraphics[width=\linewidth]{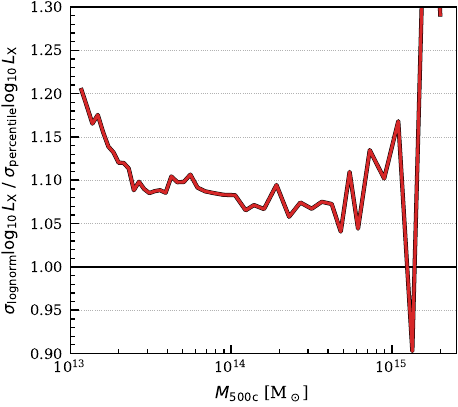}
    \caption{Ratio between the scatter measured using the percentile method ($P_{84} - P_{16}$), and the lognormal method as a function of halo mass. The lognormal method over-estimates the scatter by at least 5\% across the entire mass range, increasing to 20\% for the lowest masses ($10^{13}~\mathrm{M_\odot}$).}
    \label{fig:lognormal_scatter_comparison}
\end{figure}

In Fig.~\ref{fig:lognormal_scatter_comparison} we show that assuming a lognormal distribution over-estimates the scatter by at least 5\% at all masses. For the lowest masses, $M_{\rm 500c} < 10^{13.5}~\mathrm{M_\odot}$, the difference increases and the lognormal method can over-estimate the scatter by up to 20\%.

\section{Best fitting parameter values for redshift scatter}
This appendix lists the best fitting parameters for the broken power law describing the scatter in $L_{\rm X}$--$M_{\rm 500c}$ and the redshift evolution of those parameters. The broken power law and its evolution are shown in Fig.~\ref{fig:scatter_redshift_FLAMINGO_TNG}. Table~\ref{tab:best_fit_scatter_redshift_params_tng} has the best fitting parameters for the broken power law fit to results from the TNG300 + TNG-Cluster simulations.
\begin{table}
    \centering
    \caption{Best fitting parameter values to the redshift evolution (Eq.~\ref{eq:redshift_evolution_broken_powerlaw}) of the broken power law (Eq.  \ref{eq:broken_powerlaw_scatter}) from TNG300+TNG+Cluster.}
    \begin{tabular}{l|r|r|r} \hline
         & $\beta_1$ & $\beta_2$ & $\beta_3$ \\ \hline
       norm  & -0.0034 & 0.0424 & 0.2307\\
       slope 1 &  -0.0546 & 0.0550 & -0.2771\\
       slope 2 & 0.0350 & -0.2067 & 0.1461\\
       break-point & -0.0665 & 0.02376 & 13.5770\\ \hline
    \end{tabular}
    \label{tab:best_fit_scatter_redshift_params_tng}
\end{table}

\section{Best fitting parameter values for scatter reduction}
This appendix lists the best fitting parameter values for Equation~\ref{eq:fit_second_property_log_mass}, which can be used to reduce the scatter of the X-ray luminosity -- halo mass relation in different FLAMINGO feedback variations, and at different redshifts for the fiducial model. The parameters for $z=0$ in the fiducial L2p8 simulation are in Table~\ref{tab:l2p8_params}, for the $f_{\rm gas}+2\sigma$ model in Table~\ref{tab:fgas_p2_params}, and for the $f_{\rm gas}-8\sigma$ model in Table~\ref{tab:fgas_m8_params}. The parameters for higher redshifts of the L2p8 simulation are, for $z=0.5$ in Table~\ref{tab:l2p8_z05_params}, and for $z=1$ in Table~\ref{tab:l2p8_z10_params}.

\begin{table*}
\centering
\caption{Best fitting parameter values for equation \ref{eq:fit_second_property_log_mass} in L2p8.}
\begin{tabular}{lrrrrrrrr}\hline
$\mathcal{Y}$ & $c_1|_\alpha$ & $c_2|_\alpha$ & $c_3|_\alpha$ & $c_4|_\alpha$ & $c_1|_\beta$ & $c_2|_\beta$ & $c_3|_\beta$ & $c_4|_\beta$ \\ \hline
$f_{\rm gas}$ & 20.81 & -4.60 & 0.34 & -0.01 & 3348.90 & -727.49 & 52.58 & -1.26\\ 
Y & -30.78 & 6.34 & -0.43 & 0.01 & -1222.22 & 250.07 & -17.00 & 0.38\\ 
$E_{\rm kin} / E_{\rm therm}$ & -11.64 & 2.43 & -0.17 & 0.00 & 877.43 & -184.12 & 12.85 & -0.30\\ 
$T_{\rm gas}$ & -27.82 & 5.87 & -0.41 & 0.01 & 7837.13 & -1631.94 & 112.99 & -2.60\\ 
$t_*$ & -18.51 & 3.83 & -0.26 & 0.01 & -5275.32 & 1077.48 & -73.20 & 1.65\\ 
$M_{\rm MMBH}$ & -34.11 & 7.07 & -0.49 & 0.01 & -1460.87 & 307.95 & -21.61 & 0.50\\ 
$f_{\rm sat}$ & -22.10 & 4.69 & -0.33 & 0.01 & 283.69 & -58.35 & 3.99 & -0.09\\ 
$c_{\rm DM}$ & -48.31 & 10.30 & -0.73 & 0.02 & -355.54 & 74.08 & -5.15 & 0.12\\ 
\hline
\end{tabular} \label{tab:l2p8_params}
\end{table*}

\begin{table*}
\centering
\caption{Best fitting parameter values for equation \ref{eq:fit_second_property_log_mass} in $f_{\rm gas}+2\sigma$.}
\begin{tabular}{lrrrrrrrr}\hline
$\mathcal{Y}$ & $c_1|_\alpha$ & $c_2|_\alpha$ & $c_3|_\alpha$ & $c_4|_\alpha$ & $c_1|_\beta$ & $c_2|_\beta$ & $c_3|_\beta$ & $c_4|_\beta$ \\ \hline
$f_{\rm gas}$ & -68.06 & 14.52 & -1.03 & 0.02 & 2331.36 & -518.08 & 38.29 & -0.94\\ 
Y & -102.66 & 21.89 & -1.55 & 0.04 & -3193.40 & 670.75 & -46.85 & 1.09\\ 
$E_{\rm kin} / E_{\rm therm}$ & -64.54 & 13.81 & -0.98 & 0.02 & 1410.00 & -303.08 & 21.67 & -0.52\\ 
$T_{\rm gas}$ & -77.42 & 16.56 & -1.18 & 0.03 & 18681.13 & -3967.98 & 280.45 & -6.60\\ 
$t_*$ & -67.95 & 14.43 & -1.02 & 0.02 & -14365.35 & 3035.42 & -213.53 & 5.00\\ 
$M_{\rm MMBH}$ & -115.61 & 24.70 & -1.76 & 0.04 & -2077.87 & 447.48 & -32.06 & 0.76\\ 
$f_{\rm sat}$ & -73.08 & 15.61 & -1.11 & 0.03 & 388.99 & -82.31 & 5.79 & -0.14\\ 
$c_{\rm DM}$ & -70.47 & 15.03 & -1.07 & 0.03 & -493.01 & 102.72 & -7.12 & 0.16\\ 
\hline
\end{tabular} \label{tab:fgas_p2_params}
\end{table*}

\begin{table*}
\centering
\caption{Best fitting parameter values for equation \ref{eq:fit_second_property_log_mass} in $f_{\rm gas}-8\sigma$.}
\begin{tabular}{lrrrrrrrr}\hline
$\mathcal{Y}$ & $c_1|_\alpha$ & $c_2|_\alpha$ & $c_3|_\alpha$ & $c_4|_\alpha$ & $c_1|_\beta$ & $c_2|_\beta$ & $c_3|_\beta$ & $c_4|_\beta$ \\ \hline
$f_{\rm gas}$ & 58.86 & -12.50 & 0.88 & -0.02 & 308.33 & -58.20 & 3.55 & -0.07\\ 
Y & 10.11 & -2.07 & 0.14 & -0.00 & 745.50 & -162.79 & 11.81 & -0.28\\ 
$E_{\rm kin} / E_{\rm therm}$ & 24.10 & -5.18 & 0.37 & -0.01 & -252.92 & 59.44 & -4.61 & 0.12\\ 
$T_{\rm gas}$ & 15.10 & -3.13 & 0.22 & -0.00 & -10977.10 & 2368.32 & -170.10 & 4.07\\ 
$t_*$ & 38.92 & -8.42 & 0.61 & -0.01 & 7493.29 & -1617.26 & 115.97 & -2.76\\ 
$M_{\rm MMBH}$ & 33.69 & -7.42 & 0.54 & -0.01 & -206.23 & 39.25 & -2.47 & 0.05\\ 
$f_{\rm sat}$ & 40.86 & -8.86 & 0.64 & -0.02 & -340.18 & 74.53 & -5.43 & 0.13\\ 
$c_{\rm DM}$ & 67.55 & -14.66 & 1.06 & -0.03 & 544.17 & -115.21 & 8.10 & -0.19\\ 
\hline
\end{tabular}\label{tab:fgas_m8_params}
\end{table*}

\begin{table*}
\centering
\caption{Best fitting parameter values for equation \ref{eq:fit_second_property_log_mass} in L2p8 $z = 0.5$.}
\begin{tabular}{lrrrrrrrr}\hline
$\mathcal{Y}$ & $c_1|_\alpha$ & $c_2|_\alpha$ & $c_3|_\alpha$ & $c_4|_\alpha$ & $c_1|_\beta$ & $c_2|_\beta$ & $c_3|_\beta$ & $c_4|_\beta$ \\ \hline
$f_{\rm gas}$ & 12.17 & -2.57 & 0.18 & -0.00 & 608.32 & -135.62 & 10.04 & -0.25\\ 
Y & 11.42 & -2.62 & 0.20 & -0.00 & -963.72 & 197.64 & -13.45 & 0.30\\ 
$E_{\rm kin} / E_{\rm therm}$ & 19.90 & -4.24 & 0.30 & -0.01 & 1562.13 & -330.96 & 23.33 & -0.55\\ 
$T_{\rm gas}$ & -15.37 & 3.27 & -0.23 & 0.01 & 2736.53 & -564.13 & 38.62 & -0.88\\ 
$t_*$ & -13.16 & 2.78 & -0.20 & 0.00 & -4743.84 & 997.13 & -69.83 & 1.63\\ 
$M_{\rm MMBH}$ & -47.36 & 9.95 & -0.69 & 0.02 & -2014.50 & 428.43 & -30.33 & 0.71\\ 
$f_{\rm sat}$ & 6.56 & -1.37 & 0.09 & -0.00 & 661.70 & -139.97 & 9.85 & -0.23\\ 
$c_{\rm DM}$ & -19.73 & 4.19 & -0.30 & 0.01 & -350.25 & 74.87 & -5.34 & 0.13\\ 
\hline
\end{tabular} \label{tab:l2p8_z05_params}
\end{table*}

\begin{table*}
\centering
\caption{Best fitting parameter values for equation \ref{eq:fit_second_property_log_mass} in L2p8 $z = 1$.}
\begin{tabular}{lrrrrrrrr}\hline
$\mathcal{Y}$ & $c_1|_\alpha$ & $c_2|_\alpha$ & $c_3|_\alpha$ & $c_4|_\alpha$ & $c_1|_\beta$ & $c_2|_\beta$ & $c_3|_\beta$ & $c_4|_\beta$ \\ \hline
$f_{\rm gas}$ & 8.98 & -1.89 & 0.13 & -0.00 & -6751.63 & 1469.17 & -106.48 & 2.57\\ 
Y & 11.07 & -2.38 & 0.17 & -0.00 & -3500.45 & 758.79 & -54.78 & 1.32\\ 
$E_{\rm kin} / E_{\rm therm}$ & 51.33 & -11.02 & 0.79 & -0.02 & 2853.04 & -614.69 & 44.10 & -1.05\\ 
$T_{\rm gas}$ & 15.72 & -3.48 & 0.26 & -0.01 & -1156.88 & 268.25 & -20.65 & 0.53\\ 
$t_*$ & -26.94 & 5.83 & -0.42 & 0.01 & -4983.92 & 1067.75 & -76.26 & 1.82\\ 
$M_{\rm MMBH}$ & -35.34 & 7.43 & -0.52 & 0.01 & -2461.13 & 529.94 & -37.99 & 0.91\\ 
$f_{\rm sat}$ & 100.59 & -21.89 & 1.59 & -0.04 & 1353.53 & -292.33 & 21.02 & -0.50\\ 
$c_{\rm DM}$ & 33.00 & -7.27 & 0.53 & -0.01 & 1075.88 & -234.89 & 17.07 & -0.41\\ 
\hline
\end{tabular} \label{tab:l2p8_z10_params}
\end{table*}


\bsp	
\label{lastpage}
\end{document}